\documentclass[a4paper,11pt]{article}
\usepackage{pos}
\usepackage[utf8]{inputenc}
\usepackage[italian,english]{babel}
\usepackage{braket}
\usepackage{fancyhdr}
\usepackage{lineno}

\usepackage{tikz-cd}

\def\tor{\leftrightarrow}

\usepackage{subfigure}

\newcommand{\nn}{\nonumber\\}
\newcommand{\lsim}{\mathrel{\mathop{\kern 0pt \rlap
  {\raise.2ex\hbox{$<$}}}
  \lower.9ex\hbox{\kern-.190em $\sim$}}}
\newcommand{\gsim}{\mathrel{\mathop{\kern 0pt \rlap
  {\raise.2ex\hbox{$>$}}}
  \lower.9ex\hbox{\kern-.190em $\sim$}}}

\newcommand{\be}{\begin{equation}}
\newcommand{\ee}{\end{equation}}
\newcommand{\beqa}{\begin{eqnarray}}
\newcommand{\eeqa}{\end{eqnarray}}
\newcommand{\bea}{\begin{eqnarray}}
\newcommand{\eea}{\end{eqnarray}}

\newcommand{\beq}{\begin{equation}}
\newcommand{\eeq}{\end{equation}}

\newcommand{\sm}{\mathcal{S}}

\usepackage{pstricks}
\usepackage{color}
\usepackage{multirow}

\let\a=\alpha   \let\b=\beta      \let\d=\delta
         
      \let\l=\lambda  \let\m=\mu
\let\n=\nu

\let\d=\delta

\newcommand{\rra}{\rangle }
\newcommand{\lla}{\langle}

\newcommand{\id}{\mbox{1 \kern-.59em {\rm l}}}

\title{Nonlocal Gravity, Dark Energy and Conformal Symmetry:\\ Testing the Hierarchies of Anomaly-Induced Actions}
\ShortTitle{Nonlocal Gravity, Dark Energy and Conformal Symmetry}

\author*[a,b]{Claudio Corian\`o}
\emailAdd{claudio.coriano@le.infn.it}

\author[a]{Stefano Lionetti}
\author[a]{Matteo Maria Maglio}
\author[a]{Riccardo Tommasi}

\affiliation[a]{Dipartimento di Matematica e Fisica, Universit\`{a} del Salento 
	and INFN Sezione di Lecce, Via Arnesano 73100 Lecce, Italy and
	National Center for HPC, Big Data and Quantum Computing}
\affiliation[b]{ Institute of Nanotechnology,  National Research Council (CNR-NANOTEC), Lecce 73100, Italy}

\abstract{Conformal back-reaction generates cosmological models where the trace anomaly reflects the breaking of Weyl invariance. Analyzing these actions yields a dynamic approach to dark energy through anomaly-induced actions (AIAs), that are variational solutions of the trace anomaly functional constraint. Expanded around Minkowski space, they produce semiclassical correlators subject to hierarchical anomalous Ward identities, tied to conformal symmetry and diffeomorphism invariance. We focus on comparing the hierarchy of a specific 4-point function, particularly the 2-gravitons-2-photons correlator $(TTJJ)$, generated by AIAs, to free field theory realizations of the same correlator. We observe that the free field theory original hierarchy splits into one ordinary and one anomalous hierarchy, both satisfying the conservation Ward identities from diffeomorphism invariance. However, we find that the anomalous hierarchy derived from ordinary AIAs in both the Riegert or Fradkin-Vilkovisky gauges, are either affected by double poles or violate the hierarchy of the trace Ward identity, respectively. We show that correct forms of the anomalous hierarchies of 4-point functions (for the $TTTT$ and $TTJJ$), identified in a perturbative free field theory expansion around flat space, are characterised by anomaly poles, corresponding to a curvature expansion in $R\Box^{-1}$, together with Weyl invariant terms. We derive the effective action that generates the correct form of the hierarchy for the $TTJJ$.}

\FullConference{Corfu Summer Institute 2023 "School and Workshops on Elementary Particle Physics and Gravity" (CORFU2023),
 23 April - 6 May , and 27 August - 1 October, 2023
Corfu, Greece}

\begin{document}
\maketitle
\section{Introduction}
Conformal symmetry and its back-reaction on gravity is expected to play an important role in the physics of the early universe. It has been discussed in the context of General Relativity and semiclassical gravity in several works (see for instance \cite{Antoniadis:2006wq,Lucat:2018slu,Pelinson:2010yr,Ghosh:2020qsx}).
Indeed, modifications of General Relativity derived by integrating out conformal sectors in an external gravitational metric induce a conformal back-reaction on the same metric via the trace anomaly. 
These corrections take the form of nonlocal actions that can be tested in their consistency, as we are going to show, using the formalism of Conformal Ward identities (CWIs) in perturbation theory, by free field theory realizations, order by order in the gravitational fluctuations around flat space. Such nonlocal actions allow to address the dark energy problem in a dynamical way,  very differently from the minimal approach implemented in the standard cosmological $\Lambda$CDM model. This is phenomenologically characterised by the inclusion of a cosmological constant $(\Lambda)$ in the Einstein-Hilbert action, which is the simplest answer to the late time acceleration of our universe, but it underscores a huge hierarchy problem, according to current data. Future experiments may find out that $\Lambda$, after all, is not a constant, opening the way to alternative solutions. However, quantum anomalies may come to a rescue.\\ 
In this work, we aim to provide a brief overview of recent findings regarding the consistency of 
anomaly-induced actions (AIAs). These actions are expected to capture the anomalous content and are derived as variational solutions of the anomaly constraint. As such, they play a key role in the study of a conformal back-reaction. We subject these AIAs to a rigorous test by examining the hierarchy of equations for semiclassical correlators obtained when they are expanded around flat space. \\
Our analysis focuses on a specific 4-point function whose structure has been thoroughly investigated. Through a free field theory approach, we demonstrate that diffeomorphism invariance in flat space proves the presence of  Weyl-invariant terms in the anomalous hierarchy. These contributions are accompanied by the 
ordinary "anomaly pole parts", formerly identified in 3-point functions (e.g. in the $TTT$ and $TJJ$), connected with the presence of $R\Box^{-1}$ contributions in the anomaly effective action. We are going to illustrate this point in Section \ref{anompole}.\\
On the variational side, we investigate two forms of the nonlocal action, computed in the Riegert (R) and Fradkin-Vilkovisky (FV) gauges respectively, showing that both actions do not reproduce the expected hierarchy found perturbatively for the same correlator (the $TTJJ$). We focus on two hierarchies, the one induced by diffeomorphysim invariance and the one generated by the trace anomaly constraint. \\
We are going to show the presence of double poles in the expansion of the  AIA in the R-gauge, while in the FV-gauge the trace Ward identity is manifestly violated. The presence of Weyl-invariant terms in the perturbative expansion and the decomposition of the equations into two sub-hierarchies, was already demonstrated in \cite{Coriano:2021nvn} through the analysis of the 4-graviton vertex ($4T$), in a free field theory investigation. Such terms are absent in 3-point functions. Indeed, two previous analyses demonstrated agreement between the perturbative description of the $TTT$ correlator \cite{Coriano:2018bsy} and the prediction for the anomaly part of the same correlator computed using AIAs, as derived in the R-gauge \cite{Coriano:2017mux}, as well as in the FV-gauge. This agreement stops at the level of 3-point functions.
The method that we apply relies on recent progress on the analysis of CFT in momentum space for tensor correlators for 3-point functions. The method, introduced in \cite{Bzowski:2013sza}, has been extended to free-field theories of 4-point functions in \cite{Coriano:2021nvn,Coriano:2022jkn} (see also \cite{Coriano:2020ees}) and, more recently, to the parity-odd case \cite{Marotta:2022jrp} for non-conserved currents. In \cite{Coriano:2023hts} anomalous CP-odd correlators of chiral currents $(J_5)$ with background gauge fields and gravitons, have been reconstructed by solving the conformal constraints by the inclusion of an anomaly pole in their longitudinal chiral sector.

\subsection{Anomalies, dark matter and dark energy}
Chiral and conformal anomalies stand as potentially pivotal factors in unraveling the mysteries of dark energy and dark matter within our universe. \\
Consider, for instance, the axion, currently one of the leading contenders for dark matter. Its significance is underscored by its intricate association with a chiral anomaly inherent in the Peccei-Quinn solution of the strong CP problem.
Likewise, the tantalizing prospect that the conformal anomaly might play a role in addressing the origin of dark energy is hinted at by the cosmological constant itself. This constant, being a component of the trace of the stress-energy tensor, suggests a deeper interplay between the fundamental aspects of spacetime geometry and the dominant energy content that drives the evolution of the universe.\\
The trace anomaly extends this picture by rendering the origin of the trace of the gravitational stress tensor dynamical, and relating it to conformal symmetry, rather than leaving it to a small constant value. \\
Other issues, such as the asymmetry between matter and antimatter, as well as the origin of the cosmological magnetic fields, may also be related to other anomalies.  \\
Therefore, the idea that the dynamics of our 
universe be strongly associated with various types of quantum anomalies, is not an extremal point of view, but simply recognizes the fact that some key steps guiding its evolution are characterised by the breaking of some quantum symmetries.  Other symmetries, those related to gauge anomalies, are preserved, as indicated by the gauge stucture of the Standard Model. \\   
The investigation of these ideas from the grounds up, in order to make contact with phenomenology, may follow a simplified path, such as the investigation of effective actions of various forms, and relying on a methodology that is as close as possible to ordinary quantum field theory, leaving gravity purely classical and treated as a background. This corresponds to a semiclassical gravity approximation, but it allows to uncover some key aspects of the cosmological evolution that could be tested in the near future. It also recognizes the fundamental role played by conformal symmetry in the early stage of the cosmological evolution, no matter how simplified this approach might be. 
\section{Anomaly-induced actions }
 Conformal anomalies possess distinct features that render them more intricate compared to chiral anomalies. This complexity arises from the inclusion of both topological and non-topological terms in the anomaly functional. We recall that trace anomalies are linked with the emergence of a parity-even anomaly functional, represented by
\begin{equation}
\label{tracex1}
g_{\mu\nu}\langle T^{\mu\nu}\rangle =  b_1\, {E}_4+b_2 \, C^{\mu \nu \rho \sigma} C_{\mu \nu \rho \sigma}+b_3\, \nabla^2 R+b_4\, F^{\mu\nu}F_{\mu\nu},
\end{equation}
where $C_{\mu \nu \rho \sigma}$ denotes the Weyl tensor and ${E}_4$ stands for the Gauss-Bonnet term:
\begin{equation}
\label{gb}
	\begin{aligned}
		C^{\mu \nu \rho \sigma} C_{\mu \nu \rho \sigma} & =R^{\mu \nu \rho \sigma} R_{\mu \nu \rho \sigma}-2 R^{\mu \nu} R_{\mu \nu}+\frac{1}{3} R^2, \\
		{E}_4 & 
		\equiv E=R^{\mu \nu \rho \sigma} R_{\mu \nu \rho \sigma}-4 R^{\mu \nu} R_{\mu \nu}+R^2.
	\end{aligned}
\end{equation}
However, it was discovered by Christensen and Duff \cite{Capper:1974ic,Christensen:1978gi,Christensen:1978md,Duff:2020dqb} on dimensional grounds and through the requirement of covariance, that the structure of the trace anomaly in four dimensions can be more general than \eqref{tracex1} and constrained to the form
\begin{equation}
\label{ann}
	\mathcal{A}= b_1\, {E}_4+b_2 \, C^{\mu \nu \rho \sigma} C_{\mu \nu \rho \sigma}+b_3\, \nabla^2 R+b_4\, F^{\mu\nu}F_{\mu\nu}+f_1\, \varepsilon^{\mu \nu \rho \sigma} R_{\alpha \beta \mu \nu} R_{\,\,\,\,\, \rho \sigma}^{\alpha \beta}+f_2\,\varepsilon^{\mu\nu\rho\sigma}F_{\mu\nu}F_{\rho\sigma},
\end{equation}
which encompasses both parity-even and parity-odd terms. The parity-odd terms were explored in the action in \cite{Deser:1980kc} as potential sources of CP violation induced by gravity, offering a potential resolution to a longstanding issue. Furthermore, their relationship with anomalies was central to the study of the quantum inequivalence of different representations of antisymmetric tensor fields coupled to gravity \cite{Duff:1980qv}, \cite{Sezgin:1980tp}. Therefore, there are compelling reasons to investigate their significance within a clear physical context, particularly in the early universe, where conformal symmetry is anticipated to be fundamental.\\
Under parity inversion, all terms in the trace anomaly remain invariant except for the last two. The parity of the $F\tilde F$ term is odd, while for the $R\tilde{R}$ term, it is not naturally defined in a curved background. However, for both terms, we expect the coefficients to be real to maintain unitarity. It is worth noting that $F\tilde{F}\sim E\cdot B$ is CP-odd as well as time reversal odd. Indeed, if we regard the stress-energy tensor as a fundamental composite operator of the Standard Model (SM), the presence of imaginary coefficients would jeopardize the theory's consistency. \\
All coefficients in equation \eqref{tracex1} have been computed in the parity-even case, and their values strictly depend on the number and types of massless fields contributing to the perturbative quantum corrections, yet they are real. The parity-odd case \cite{Coriano:2023hts,Coriano:2023gxa,Coriano:2023cvf} has been discussed in the context of CFT in momentum space in some recent works. In the following we are going to limit our discussion only to the parity-even case.\\
\subsection{The effective action in the parity-even sector}
\begin{figure}[t]
	\centering
	\includegraphics[scale=0.55]{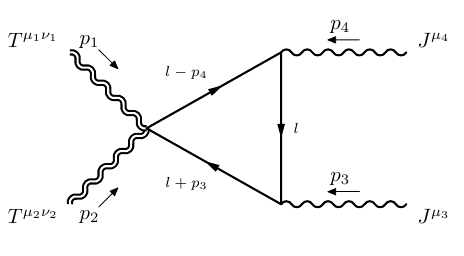}\hspace{5ex}
	\includegraphics[scale=0.55]{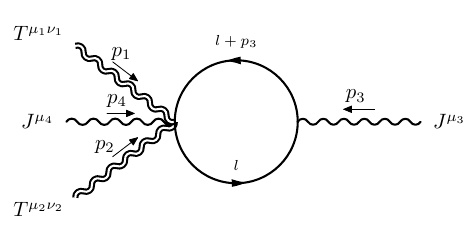}\hspace{5ex}
	\includegraphics[scale=0.55]{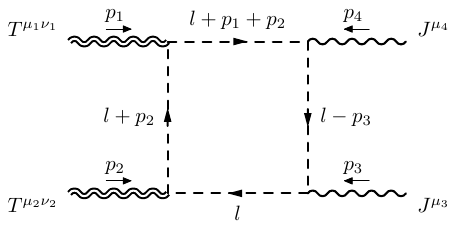}\hspace{5ex}
	\includegraphics[scale=0.55]{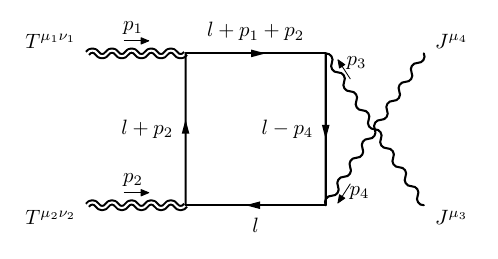}\hspace{5ex}
		\includegraphics[scale=0.45]{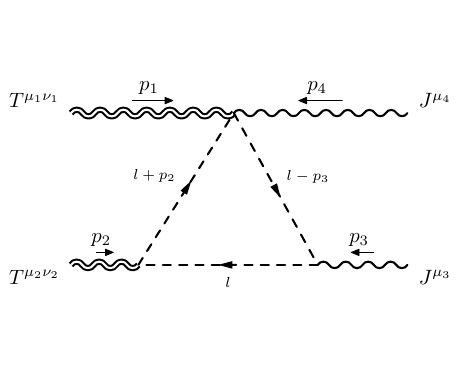}\hspace{5ex}
\caption{Examples diagrams in the perturbative realization of the $TTJJ$ with virtual fermions and scalars.\label{TriangleDiag}}
\end{figure}

In this work we will focus on the parity-even sector. We look for an effective action that can account for \eqref{tracex1} from the flat spacetime limit with $g_{\mu\nu}=\delta_{\mu\nu} + \delta g_{\mu\nu}$, in the Euclidean case

 \begin{align}
\label{exps2}
\sm(g)_B &\equiv\sm(\bar{g})_B+\sum_{n=1}^\infty \frac{1}{2^n n!} \int d^d x_1\ldots d^d x_n \sqrt{-g_1}\ldots \sqrt{-g_n}\,\langle T^{\mu_1\nu_1}\ldots \,T^{\mu_n\nu_n}\rangle_{\bar{g} B}\delta g_{\mu_1\nu_1}(x_1)\ldots \delta g_{\mu_n\nu_n}(x_n) \nn
& + \textrm{(gravity/photons terms)}
\end{align}
expressed in terms of correlation functions of multi-graviton vertices 
$\langle T^{\mu_1\nu_1}\ldots \,T^{\mu_n\nu_n}\rangle$  defined in a generic metric backgound $\bar{g}$. They are evaluated by introducing a bare $(B)$ Lagrangian containing an arbitrary  number of scalars, spin-1 and fermion fields. Other contributions, such as the mixed photon/graviton vertices, have beeen not been explicitly displayed in \eqref{exps2}. \\
The bare effective action is renormalized by the inclusion of ordinary counterterms. These are proportional to the Gauss-Bonnet term, the Weyl tensor squared and 
the square of the field strength $F^2$. The renormalized effective action satisfies anomalous conformal Ward identities which are hierarchical. \\
Diagrammatically, the pure gravitational sector is identified, in free field theory realizations, by an infinite sum of 1-loop diagrams labelled by an arbitrary number of external graviton lines. 
The mixed sector, on the other end, will include contributions such as the $TTJ$, $TJJ$ and the $TTJJ$ correlation functions. 
Associated with this correlation function, that we will study in detail in free field theory, are hierarchical costraints that take the form of CWIs (dilatation and special conformal) together with those induced by diffeomorphsm invariance of the partition function. We are going to show that the hierarchy associated with these second constraints, are not respected by the correlators predicted by AIAs in flat space. \\
The $TTJJ$ is the first correlation function that allows to test the consistencly between the conformal anomaly effective action and the conformal constraints that it has to satisfy.  \\
We are going to summarize the way this test is performed and identify the missing contributions that are necessay in order to relate the two approaches, the perturbative one and the variational one.  
The diagrammatic expansion of the $\braket{TTJJ}$ for the fermionic and scalar cases are given  in Fig.\ref{TriangleDiag}.\\
In Dimensional Regularization (DR) the renormalized effective action is defined by the inclusion of three counterterms 
\begin{equation}
	{Z}_R(g)=\, \mathcal{N}\int D\Phi\,e^{-S_0(g,\chi) + \frac{b' }{\epsilon}V_E(g,d) +  \frac{b}{\epsilon}V_{C^2}(g,d) - S_{count}(g,A)},
\end{equation} 
where $\mathcal{N}$ is a normalization constant, $\epsilon=d-4$, and 
\begin{equation}
\begin{split}
	V_{C^2}(g, d)\equiv & \mu^{\varepsilon}\int\,d^dx\,\sqrt{-g}\, C^2,  \\
	V_{E}(g,d)\equiv &\mu^{\varepsilon} \int\,d^dx\,\sqrt{-g}\,E , 
	\end{split}	\label{ffr}
\end{equation}

In order to remove these divergences of the mixed graviton/gauge correlators we add to the action the counterterm 
\begin{align}
	S_{count}(g,A)\equiv -\frac{1}{\varepsilon}V_{F^2}(g, d)\equiv -\frac{\mu^{\varepsilon}}{\varepsilon}\sum_{I=f,s}\,n_I\int d^dx\,\sqrt{-g}\left(\beta_c(I)\,F^2\right),
\end{align}
corresponding to the field strength $F^2=F^{\mu\nu}F_{\mu\nu}$ where the coefficients $\beta_c(I)$ refer to the scalar and fermion contributions. The renormalized actions and vertices are extracted, ordinarily, from the enormalized effective action
\beq
S_{R}=S_0 +S_{count}
\eeq
where the counterterms are purely gravitational and/or contain gauge fields. 
In the free field theory realization, which allows to define a consistent framework for the analysis of the CWIs, the gauge constraints and diffeomorphism invariance in the curved background, one can  move a long way in the analysis of such one-loop contributions to the effective action. 
The conservation/anomalous contraints - on the currents and the stress energy tensors - by functional differentiation of $\sm(\bar{g})_R$ can be extended to any order $n$ of the expansion in the fluctuations. The CWIs take also a general form for the entire hierarchy. 
\subsection{The TTJJ in free field theory}
We will focus our attention on the case $n=4$ and, specifically, on the $TTJJ$ correlator, which is more manageable compared to the $4T$, for being of lower tensor rank. The reason for moving to 4-point functions is that the disagreement between the inconsistency of the variational solutions of the trace anomaly contraints start to emerge at this order. 
 The analysis is performed by extending a method of investigation of the CWIs developed in the context of CFT in momentum space. It is well-known that for 3-point functions the conformal constraints are powerful enough to fix the structure of such correlators, modulo few constants that depend on the spacetime dimensions. Higher point functions need to be bootstrapped, an approach which, so far, has been investigated mostly in coordinate space. The approach is based on the operator product expansion, and for this reason it is rather inefficient as far as the analysis of the anomaly contributions are concerned, since these come from regions where all the external points of the correlator coalesce in coordinate space.\\
 Chiral and conformal anomaly contributions are easily captured in perturbation theory using the usual diagrammatic Feynman expansion, arrested at one-loop. We allow free conformal field theories of scalars, spin-1 and fermions to run in the corrections. In $d=4$ spacetime dimensions the general structure of the 3-point functions of correlators of stress energy tensors and/or/photons is reproduced by a superposition of these three conformal sectors, each with arbitrary multiplicities.  \\
 This provides the simplest realization of the anomaly effective action, expanded in the external fluctuations around flat space and in the presence of background photons.
The structure of the renormalized constraints, their hierarchical nature and the organization of the hierarchy in terms of anomalous and non anomalous parts can be worked out explicitly. \\
We will review the case of the $TTJJ$, a 4-point functions of gravitons and gauge fields, that allows us to uncover the limitations of anomaly induced actions, solutions of \eqref{tracex1}.\\
The $TTJJ$ correlator around flat space is extracted by taking four derivatives of the bare effective action $\sm_B$ with respect to the metric and the gauge field, evaluated when the sources are turned off 
	\begin{align}
	\braket{T^{\mu_1\nu_1}(x_1)\,T^{\mu_2\nu_2}(x_2)\,J^{\mu_3}(x_3)\,J^{\mu_4}(x_4)}=\,4\,\frac{\delta^4\,\mathcal{S}_B}{\delta g_{\mu_1\nu_1}(x_1)\,\delta g_{\mu_2\nu_2}(x_2)\,\delta A_{\mu_3}(x_3)\,\delta A_{\mu_4}(x_4)}\Bigg|_{g=\delta,\,A=0}\,.
	\end{align}

The correlator is part of the general conformal anomaly effective action in the presence both of a metric and of abelian gauge fields. The perturbative CWIs are organized in a way that allows to identify two folds in the general hierarchy of the equations. Both folds/hierarchies are identified by expanding the conformal constraints of the partition function order by order in the metric and background gauge field fluctuations. \\
The symmetry constraints on $\sm_R$, induced by the coefficients of the expansion \eqref{exps2}, take the form of hierarchical Ward Identities (WIs). The conformal constraints, for instance, are linked to the Weyl invariance of the renormalized effective action and its breaking. The derivation of the corresponding WIs can be performed directly from either $\sm_B$ or $\sm_R$, as demonstrated in \cite{Coriano:2017mux,Coriano:2021nvn}.\\
Recall that in a curved background, for a certain action $\sm(g)$, Weyl invariance is expressed as a symmetry of the form 
\begin{equation}
\sm(g)=\sm(\bar{g}) \quad \textrm{when} \quad g_{\mu\nu}=\bar{g}_{\mu\nu} e^{2 \phi}.
\end{equation}
The relation between $g$ and $\bar{g}$ defines a conformal decomposition, which remains valid under the gauge transformation  
\begin{equation}
\bar g \to \bar g\,e^{2 \sigma}, \quad \phi\to \phi - \sigma,
\end{equation}
where $\sigma(x)$ is a local shift. The renormalization of the quantum corrections, via the counterterms above, breaks this symmetry. In the case of a flat background, one is essentially performing the $\phi\to 0$ limit of $\sm_R$ after performing the metric variations, with the dilaton variation $\frac{\delta}{\delta \phi}$ replaced by $2 g_{\mu\nu}\frac{\delta}{\delta g_{\mu\nu}}$. \\
In general, on the bare functional $\sm_B(g)$, one derives the relation 
\begin{equation}
\label{anomx}
\frac{\delta \sm_B}{\delta \phi(x)}=\sqrt{-g} \,g_{\mu\nu}\,\langle T^{\mu\nu}\rangle, 
\end{equation} 
and its invariance under Weyl transformations  
\begin{equation}
\label{www}
\delta_\phi g_{\mu\nu}= 2  g_{\mu\nu} \delta\phi,
\end{equation}
and diffeomorphisms   
\begin{equation} 
\delta_\epsilon g_{\mu\nu}=-\nabla_{\mu}\epsilon_{\nu}- \nabla_{\nu}\epsilon_{\mu}, 
\end{equation}
are summarized by the constraints
\begin{equation} 
\label{eww}
\delta_\phi \sm_B=0, \quad \delta_\epsilon \sm_B=0,
\end{equation}
leading to trace and conservation conditions of the quantum averages of $T^{\mu\nu}$
\begin{equation}
\label{comby}
\langle T^\mu_\mu\rangle=0, \quad \nabla_\mu\langle T^{\mu\nu}\rangle=0.
\end{equation}
Ordinary trace and conservation WIs can be derived from the equations above by functional differentiations of $\sm_B(g)$ with respect to the background metric. As long as we stay away from $d=4$ and include conformal fields in the classical action $\sm_0$, we have exact CWIs derived from the condition of invariance of the generating functional $\sm_B$ with respect to diffeomorphisms and Weyl transformations. Anomalous CWIs are derived by replacing the effective action $\sm_B$ with the renormalized one $\sm_R$. \\
We proceed to discuss the derivation of the conformal and conservation Ward Identities (WIs) for the correlator, illustrating its decomposition following the approach of \cite{Bzowski:2013sza} into a transverse traceless sector, a longitudinal sector, and a trace sector. For our analysis, only the trace and conservation WIs will be relevant.\\
Assuming that the generating functional of the theory is invariant under the action of certain symmetry groups, the correlation function $\braket{TTJJ}$ satisfies
\begin{equation}
\sum_{j=1}^4\,G_g(x_j)\braket{T^{\mu_1\nu_1}(x_1)T^{\mu_2\nu_2}(x_2)J^{\mu_3}(x_3)J^{\mu_4}(x_4)}=0,
\end{equation}
where $G_g$ represents the generators of infinitesimal symmetry transformations. These constraints stem from the invariance of the generating functional under symmetry transformations:
\begin{equation}
\sm_B[g',A']=\sm_B[g,A],
\end{equation}
which can be equivalently expressed as
\begin{equation}
\int\,d^dx\left(\frac{\delta\sm_B}{\delta g_{\mu\nu}}\delta g_{\mu\nu}+\frac{\delta\sm_B}{\delta A_\mu^a}\delta A^a_\mu\right)=0.\label{invW}
\end{equation}
Among these constraints, the conservation Ward Identity (WI) in flat space of the energy-momentum tensor can be obtained by requiring the invariance of $\sm_B[g,A]$ under diffeomorphisms $x^\mu\to x^\mu+\epsilon^\mu(x)$, for which the variation of the metric and the gauge fields are the corresponding Lie derivatives. In the case of a non-Abelian $SU(N)$ gauge field $A_{\mu}^a$ ($a=1,2,\ldots, N^2-1)$, for instance, we obtain
\begin{align}
\delta A_\mu^a&=-\epsilon^\alpha\nabla_\alpha A^a_\mu-A_\alpha^a\nabla_\mu\epsilon^\alpha,\\
\delta g_{\mu\nu}&=-\nabla_\mu\epsilon_\nu-\nabla_\nu\epsilon_\mu.
\end{align}
Inserting these variations into \eqref{invW} and integrating by parts, we obtain the conservation WI:
\begin{align}
\nabla_\mu\braket{T^{\mu\nu}}+\left(\partial^\mu A^{a\nu}-\partial^\nu A^{a\mu}\right)\braket{J^a_\mu}+A^{a\nu}\,\nabla_\mu\braket{J^{a\mu}}=0.\label{cons1}
\end{align}

Similarly, the requirement of invariance under a gauge transformation with a parameter $\theta^a(x)$ gives
\begin{align}
\delta A_\mu^a&=\partial_\mu\theta^a+g\,f^{abc}A_\mu^b\theta^c,\\
\delta g_{\mu\nu}&=0.
\end{align}
The invariance of the generating functional under gauge transformations gives
\begin{align}
\nabla_\mu\braket{J^{a\mu}}=gf^{abc}A_\mu^c\braket{J^{b\,\mu}}. 
\end{align}
Inserting this equation into \eqref{cons1} we obtain the conservation WIs
\begin{subequations}
\begin{align}
\nabla_\mu\braket{T^{\mu\nu}}+F^{a\,\mu\nu}\braket{J^a_\mu}&=0,\\
\nabla_\mu\braket{J^{a\mu}}-gf^{abc}A_\mu^c\braket{J^{b\,\mu}}&=0. 
\end{align}
\end{subequations}
In the Abelian case, which is our focus, diffeomorphism and gauge invariance give the relations
\begin{subequations}
\begin{align}
&\nabla_\mu\braket{T^{\mu\nu}}+F_{\mu\nu}\braket{J^{\mu}}=0,\\
&\nabla_\mu\braket{J^{\mu}}=0. 
\end{align}\label{consWI}
\end{subequations}
\subsection{A hierarchy from diffeomorphism invariance  ($4\to 3$)}
By functional differentiation of \eqref{consWI}, we derive ordinary WIs for the various correlators involving energy-momentum tensors and conserved currents. In the $\braket{TTJJ}$ case, after a Fourier transform, we obtain the conservation equation
	\begin{align} 
	& {p_1}_{\mu_1} \braket{T^{\mu_1 \nu_1} (p_1) T^{\mu_2 \nu_2} (p_2) J^{\mu_3} (p_3) J^{\mu_4} (p_4)}  \notag\\
	& 
	=\bigg[2\ {p_2}_{\lambda_1} \delta^{\nu_1 (\mu_2} \braket{T^{\nu_2) \lambda_1} (p_1+p_2) J^{\mu_3} (p_3) J^{\mu_4} (p_4)}  - 
	{p_2}^{\nu_1}  \braket{T^{\mu_2 \nu_2} (p_1+p_2) J^{\mu_3} (p_3) J^{\mu_4} (p_4)} \bigg] \notag\\ 
	&
	+ 2 \bigg\{  \bigg[ \delta^{\nu_1 (\mu_2} p_3^{\nu_2)}  \braket{J^{\mu_3} (p_1+p_2+p_3) J^{\mu_4} (p_4)} - \delta^{\nu_1 (\mu_2} \delta^{\nu_2) \mu_3}{p_3}_{\lambda_1 } \braket{J^{\lambda_1} (p_1+p_2+p_3) J^{\mu_4} (p_4)} 
	\notag\\ 
	&\qquad + \, \frac{1}{2}\delta^{\mu_3 \nu_1} p_{3\lambda_1} \braket{J^{\lambda_1}(p_1+p_3) T^{\mu_2 \nu_2}(p_2) J^{\mu_4}(p_4) }- \frac{1}{2}p_3^{\nu_1 } \braket{J^{\mu_3}(p_1+p_3) T^{\mu_2 \nu_2}(p_2) J^{\mu_4}(p_4) } \bigg]+\bigg[ (3 \tor 4) \bigg]\bigg\} ,\label{Cons1}
\end{align}
where the notation $(3\leftrightarrow 4)$ means the exchange of the subscript $3$ with $4$, and the vector current Ward identities
	\begin{equation} 
		p_{i\,\mu_i} \braket{T^{\mu_1 \nu_1} (p_1) T^{\mu_2 \nu_2} (p_2) J^{\mu_3} (p_3)  J^{\mu_4} (p_4)} = 0, \qquad i=3,4.\label{Cons2}
	\end{equation}
	In our conventions, all the momenta, in a given correlator, are incoming.
Furthermore we consider the invariance of the generating functional under Weyl transformations for which the fields transform as in \eqref{www} and
\begin{align}
\delta_\sigma\,A_\mu^a&=0
\end{align}
giving the naive trace Ward identity 
\begin{align}
g_{\mu\nu}\braket{T^{\mu\nu}}=0. \label{traceNaive}
\end{align}
The functional differentiation of \eqref{traceNaive} gives the (non-anomalous) condition
\begin{align}
\delta_{\mu_1 \nu_1} \braket{T^{\mu_1 \nu_1} (p_1) T^{\mu_2 \nu_2} (p_2) J^{\mu_3} (p_3) J^{\mu_4} (p_4) } = -2 \, \braket{T^{\mu_2 \nu_2} (p_1+p_2) J^{\mu_3} (p_3) J^{\mu_4} (p_4) }.\label{Cons3}
\end{align}
The analysis of the 3-point functions and their WIs/CWIs, in the $TJJ$, has been investigated in previous works, and are the first stepping stones in order to proceed with the current extension \cite{Coriano:2018bbe}. Renormalization will modify the equation above. 
\section{Decomposition of the correlator}\label{decomp}
As previously noted, the $\braket{TTJJ}$ correlator can be systematically decomposed into transverse, longitudinal, and trace components \cite{Bzowski:2013sza}, leveraging their symmetries. The approach needs to be extended to 4-point functions. 
This general strategy is that of expressing the operators $T$ and $J$ in terms of their transverse traceless part and the longitudinal (local) components:
	\begin{align}
		T^{\mu_i\nu_i}(p_i)&\equiv t^{\mu_i\nu_i}(p_i)+t_{loc}^{\mu_i\nu_i}(p_i),\label{decT}\\
		J^{\mu_i}(p_i)&\equiv j^{\mu_i}(p_i)+j_{loc}^{\mu_i}(p_i),\label{decJ}
	\end{align}
	where
	\begin{align}
		\label{loct}
		t^{\mu_i\nu_i}(p_i)&=\Pi^{\mu_i\nu_i}_{\alpha_i\beta_i}(p_i)\,T^{\alpha_i \beta_i}(p_i), &&t_{loc}^{\mu_i\nu_i}(p_i)=\Sigma^{\mu_i\nu_i}_{\alpha_i\beta_i}(p)\,T^{\alpha_i \beta_i}(p_i),\\
		j^{\mu_i}(p_i)&=\pi^{\mu_i}_{\alpha_i}(p_i)\,J^{\alpha_i }(p_i), &&\hspace{1ex}j_{loc}^{\mu_i}(p_i)=\frac{p_i^{\mu_i}\,p_{i\,\alpha_i}}{p_i^2}\,J^{\alpha_i}(p_i).
	\end{align}
Here, we introduce the transverse-traceless ($\Pi$), transverse $(\pi)$, and longitudinal ($\Sigma$) projectors as 
\begin{align}
	\label{prozero}
	\pi^{\mu}_{\alpha} & = \delta^{\mu}_{\alpha} - \frac{p^{\mu} p_{\alpha}}{p^2}, \\
	\Pi^{\mu \nu}_{\alpha \beta} & = \frac{1}{2} \left( \pi^{\mu}_{\alpha} \pi^{\nu}_{\beta} + \pi^{\mu}_{\beta} \pi^{\nu}_{\alpha} \right) - \frac{1}{d - 1} \pi^{\mu \nu}\pi_{\alpha \beta}\label{TTproj}, \\
	\Sigma^{\mu_i\nu_i}_{\alpha_i\beta_i}&=\frac{p_{i\,\beta_i}}{p_i^2}\Big[2\delta^{(\nu_i}_{\alpha_i}p_i^{\mu_i)}-\frac{p_{i\alpha_i}}{(d-1)}\left(\delta^{\mu_i\nu_i}+(d-2)\frac{p_i^{\mu_i}p_i^{\nu_i}}{p_i^2}\right)\Big]+\frac{\pi^{\mu_i\nu_i}(p_i)}{(d-1)}\delta_{\alpha_i\beta_i}\equiv\mathcal{I}^{\mu_i\nu_i}_{\alpha_i}p_{i\,\beta_i} +\frac{\pi^{\mu_i\nu_i}(p_i)}{(d-1)}\delta_{\alpha_i\beta_i}\label{Lproj}.
\end{align}
Utilizing these projectors, the correlator can be expressed as
\begin{align}
	&\braket{ T^{\mu_1 \nu_1} (p_1) T^{\mu_2 \nu_2} (p_2) J^{\mu_3} (p_3) J^{\mu_4} (p_4) } \notag\\[1ex]
	&\hspace{1cm}=\braket{ t^{\mu_1 \nu_1} (p_1) t^{\mu_2 \nu_2} (p_2) j^{\mu_3} (p_3) j^{\mu_4} (p_4) } +	\braket{ t^{\mu_1 \nu_1} (p_1) t^{\mu_2 \nu_2} (p_2) j^{\mu_3} (p_3) j^{\mu_4} (p_4) } _{loc}\label{decTTJJ}
\end{align}
where the transverse traceless sector satisfies the conditions
\begin{align}
\delta_{\mu_i\nu_i}\braket{ t^{\mu_1 \nu_1} (p_1) t^{\mu_2 \nu_2} (p_2) j^{\mu_3} (p_3) j^{\mu_4} (p_4) } &=0,\quad i=1,2\,,\label{tracel}\\
p_{\mu_i}\braket{ t^{\mu_1 \nu_1} (p_1) t^{\mu_2 \nu_2} (p_2) j^{\mu_3} (p_3) j^{\mu_4} (p_4) } &=0,\quad i=1,\dots,4\,\label{transverse},
\end{align}
and the contribution from the other sectors, collected in the local part, are expressed in the form 
\begin{align}
&\braket{ t^{\mu_1 \nu_1} (p_1) t^{\mu_2 \nu_2} (p_2) j^{\mu_3} (p_3) j^{\mu_4} (p_4) } _{loc}=\braket{ t_{loc}^{\mu_1 \nu_1}T^{\mu_2 \nu_2} J^{\mu_3} J^{\mu_4}  } +\braket{ T^{\mu_1 \nu_1}  t_{loc}^{\mu_2 \nu_2}  J^{\mu_3} J^{\mu_4} } -\braket{ t_{loc}^{\mu_1 \nu_1} t_{loc}^{\mu_2 \nu_2}J^{\mu_3} J^{\mu_4}}\notag\\[1ex]
&= \bigg[\left(\mathcal{I}^{\mu_1\nu_1}_{\alpha_1}p_{1\,\beta_1} +\frac{\pi^{\mu_1\nu_1}(p_1)}{(d-1)}\delta_{\alpha_1\beta_1}\right)\delta^{\mu_2}_{\alpha_2}\delta^{\nu_2}_{\beta_2}+\left(\mathcal{I}^{\mu_2\nu_2}_{\alpha_2}p_{2\,\beta_2} +\frac{\pi^{\mu_2\nu_2}(p_2)}{(d-1)}\delta_{\alpha_2\beta_2}\right)\delta^{\mu_1}_{\alpha_1}\delta^{\nu_1}_{\beta_1}\notag\\
&\hspace{1cm}-\left(\mathcal{I}^{\mu_1\nu_1}_{\alpha_1}p_{1\,\beta_1} +\frac{\pi^{\mu_1\nu_1}(p_1)}{(d-1)}\delta_{\alpha_1\beta_1}\right)\left(\mathcal{I}^{\mu_2\nu_2}_{\alpha_2}p_{2\,\beta_2} +\frac{\pi^{\mu_2\nu_2}(p_2)}{(d-1)}\delta_{\alpha_2\beta_2}\right)\bigg]\,\braket{T^{\alpha_1\beta_1}T^{\alpha_2\beta_2}J^{\mu_3}J^{\mu_4}}.\label{loc}
\end{align}
The trace and longitudinal contributions, contained in the second term of \eqref{decTTJJ} and explicitly expressed in \eqref{loc}, are subject to the conservation WIs \eqref{Cons1}, \eqref{Cons2}, and \eqref{Cons3}. Consequently, the undetermined portion of the correlator resides within its transverse-traceless ($ttjj$) sector, as the remaining longitudinal and trace contributions - the local terms - are linked to lower point functions through conservation and trace WIs. Hence, we can advance by examining the general decomposition of the transverse-traceless part $\braket{ttjj}$ into a product of form factors and tensor structures. \\
Due to conditions \eqref{tracel} and \eqref{transverse}, such a sector, utilizing the transverse and traceless projectors, assumes the form
	\begin{equation} 
		\braket{ t^{\mu_1 \nu_1} (p_1) t^{\mu_2 \nu_2} (p_2) j^{\mu_3} (p_3) j^{\mu_4} (p_4) } =\Pi^{\mu_1 \nu_1}_{\alpha_1 \beta_1} (p_1)  \Pi^{\mu_2 \nu_2}_{\alpha_2 \beta_2} (p_2) \pi^{\mu_3}_{\alpha_3} (p_3) \pi^{\mu_4}_{\alpha_4} (p_4) X^{\alpha_1 \beta_1 \alpha_2 \beta_2 \alpha_3\alpha_4},\label{ttjj}
	\end{equation}
where $X^{\alpha_1\dots\alpha_4}$ is a general rank six tensor constructed from products of metric tensors and momenta with appropriate index selection. Notably, due to the presence of the projectors in \eqref{ttjj}, the terms $\delta^{\alpha_i\beta_i}$, $i=1,2$, or $p_i^{\alpha_i}$, $i=1,\dots,4$, cannot serve as fundamental tensors and vectors for constructing the $X^{\alpha_1\dots\alpha_4}$ tensor. Additionally, the conservation of total momentum
	\begin{equation} 
		p_1^{\alpha_i} + p_2^{\alpha_i} + p_3^{\alpha_i} + p_4^{\alpha_i} = 0 ,
	\end{equation}
permits the selection of a pair of momenta for each index $\alpha_i$, to be utilized in the general construction of $X$. The parameterizaiton of $X^{\alpha_1\dots\alpha_4}$ in terms of a minimal number of form factors is performed is such a way that the momenta are treated symmetrically. This approach is efficient, reducing the number of form factors to a minimum by exploiting the presence of a single $tt$ projector for each external momentum. Concerning the tensor structures formed from metric $\delta$s, the only non-vanishing ones in $X^{\alpha_1\dots\alpha_4}$ are:
	\begin{align}
	\delta^{\alpha_1\alpha_2},\ \delta^{\alpha_1\alpha_3},\ \delta^{\alpha_1\alpha_4},\ \delta^{\alpha_2\alpha_3},\ \delta^{\alpha_2\alpha_4},\ \delta^{\alpha_3\alpha_4}\label{choicemetr}
	\end{align}
alongside similar ones obtained by exchanging $\alpha_i\leftrightarrow\beta_i$, $i=1,2$. This strategy, introduced in \cite{Bzowski:2013sza} for 3-point functions and applied to $4$-point functions in \cite{Coriano:2021nvn}, will be elaborated upon in the subsequent section to explicitly express the $X^{\alpha_1\dots\alpha_4}$ expression in terms of the minimal number of tensor structures and form factors, in general $d$ dimensions \cite{Coriano:2019nkw}.

\subsection{Orbits of the permutations\label{OrbitsSec}}
The expression for $X^{\alpha_1\dots\alpha_4}$ relies on tensor structures and form factors, exploiting the symmetry of the correlator. The $\braket{TTJJ}$ is characterised by two types of discrete symmetries related to the permutation group: it must be symmetric under the exchange of the two gravitons ($1 \tor 2$), of the two conserved $J$ currents ($3 \tor 4$), and the combination of both transformations. We denote these transformations respectively as $P_{12}$, $P_{34}$, and $P_C=P_{12}P_{34}$. Notably, $P_{12}$ exchanges the pair of indices of the two gravitons and the associated momenta, and analogously for the two currents $J$'s.   
		The tensorial structures in $X^{\alpha_1\dots\alpha_4}$ will be constructed by utilizing the metric tensors and the momenta with a symmetric expansion in terms of different sets of independent momenta. Then, in $X^{\alpha_1\dots\alpha_4}$, there are structures of four different types, depending on the number of metric tensors and momenta used to saturate the number of free indices. We consider the general terms
		\begin{equation} 
			\delta\delta\delta ,  \, \delta\delta p p ,  \, \delta p ppp , \, pppppp,  \label{tensorsectors}
		\end{equation}
		observing that these sectors do not mix when the permutation operator $P_{ij}$ is applied. This property allows us to construct the general symmetric form of each sector separately.\\
As a first step, we determine the orbits of the $P$ operators acting on the tensor structures belonging to each tensorial sector \eqref{tensorsectors}.  This can be achieved by applying all the $P$ transformations to a tensor structure and following the "trajectory" (orbits) in the sector generated by this process.  For instance, in the sector $\delta\delta p p$, we encounter the two orbits
\begin{eqnarray}
	\begin{tikzcd}
		\delta^{\alpha_2 \beta_1} \delta^{\alpha_4 \beta_2} p_3^{\alpha_1} p_1^{\alpha_3}\arrow[r, "P_{12}"] \arrow[rd, "P_C"] & \delta^{\alpha_1 \beta_2} \delta^{\alpha_4 \beta_1} p_3^{\alpha_2} p_2^{\alpha_3} \arrow[d, "P_{34}"]  \\
		\delta^{\alpha_2 \beta_1} \delta^{\alpha_3\beta_2} p_4^{\alpha_1} p_1^{\alpha_4}  \arrow[u, "P_{34}"] & \delta^{\alpha_1 \beta_2} \delta^{\alpha_3\beta_1}  p_4^{\alpha_2} p_2^{\alpha_4} \arrow[l,"P_{12}"]
	\end{tikzcd}\label{orbitex1}
\end{eqnarray}
\begin{eqnarray}
	\begin{tikzcd}
		\delta^{\alpha_1 \alpha_2}  \delta^{\beta_1 \beta_2} p_1^{\alpha_3}p_1^{\alpha_4 }\arrow[rr, "P_{12}", " P_C" '] \arrow["P_{34}"', loop, distance=2em, in=305, out=235] &  & \delta^{\alpha_1 \alpha_2}  \delta^{\beta_1 \beta_2} p_2^{\alpha_3}p_2^{\alpha_4} \arrow["P_{34}"', loop, distance=2em, in=305, out=235].
	\end{tikzcd} \label{orbitex2}
\end{eqnarray}
In this way, we decompose every sector \eqref{tensorsectors} into orbits. Every $P$ transformation acts on an orbit irreducibly, i.e. it connects every element on the orbit. The number of orbits for all the sectors equals the number of independent form factors representing the correlator. 
In fact, a representative can be selected for each orbit to which an independent form factor can then be associated. The orbit provides a visual realization of the symmetry properties of the form factors that belong to it. \\
It is now clear that the study of the orbits of the tensor structures under the permutation group, provides directly the answer about the minimal number of independent form factors that are needed in order to describe the general solution of any $4$-point correlator. This procedure can be simply generalized to higher point correlation functions involving operators of any spin. \\
Once we identify and select representative of each orbit for every sector, then the general structure of $X^{\alpha_1\dots\alpha_4}$ can be written down quite easily. In this way, we find that in $d>4$ the general form of $X^{\alpha_1\dots\alpha_4}$ related to $\braket{TTJJ}$ is written in terms of $47$ independent form factors. This number reduces significantly when $d\le4$ (see Table \ref{generalD}). \\
\begin{table}[t]
		\centering
		\begin{tabular}{|c|c|c|} \hline&&\\[-2ex]
			Sector & $\#$ of tensor structures &$\#$ of orbits \\[1ex] \hline &&\\[-2ex]
			$\delta \delta \delta$ & 3 & 2 \\[1ex]
			$\delta \delta p p$ & 38 & 13 \\[1ex]
			$\delta p p p p$ & 73  & 21 \\[1ex]
			$p p p p p p$ & 36  & 11  \\[1ex]\hline
			Total & 150 &47 \\ \hline
		\end{tabular}
	\caption{ Number of tensor structures and independent form factors in $X^{\alpha_1\dots\alpha_4}$ for the $\braket{TTJJ}$. \label{generalD}}
	\end{table}
\subsection{Degeneracies and Lovelock's identities} 
In dimensions $d\leq 5$, the composition of $X^{\alpha_1\dots \alpha_4}$ undergoes modifications due to the degeneracies present in the tensor structures \cite{Coriano:2021nvn,Edgar:2001vv, lovelock_1970, Bzowski:2017poo}. These degeneracies lead to a decrease in the count of independent form factors, resulting in a notably simplified structure for the correlator.\\
Drawing from the insights outlined in \cite{Edgar:2001vv, lovelock_1970}, it is established that every tensor in a $d$-dimensional space corresponds to a fundamental tensor identity derived by antisymmetrizing over $d+1$ indices. Specifically, consider $\mathcal{T}^{A\qquad\ b_1\dots b_l}_{\ \ a_1\dots a_k}=\mathcal{T}^{A\qquad \ [b_1\dots b_l]}_{\ \ [a_1\dots a_k]}$, representing a trace-free tensor across all of its indices, where $A$ signifies an arbitrary number of additional lower and/or upper indices. Hence,
\begin{equation}
\mathcal{T}^{A\qquad \ [b_1\dots b_l}_{\ \ [a_1\dots a_k}\ \delta_{a_{k+1}}^{b_{l+1}}\ \dots\  \delta_{a_{k+n]}}^{b_{l+n]}}=0,\label{Lovelock}
\end{equation}
where $n\ge d-k-l+1$ and $n\ge0$.
In $d=4$, the metric $\delta$ loses its independence, allowing for a new basis construction facilitated by the antisymmetric tensor $\epsilon^{\mu_1\dots\mu_4}$ and three of the four external momenta. This reduction is feasible when a correlation function in $d$ dimensions includes at least $d-1$ independent external momenta. The resultant vector $n^\mu$ is transverse to $p_1 , p_2 , p_3$, denoted by $n\cdot p_i=0$. Utilizing this new basis, termed the $n$-$p$ basis, all tensorial structures defining the correlation function can be constructed. In this basis, the metric tensor $\delta^{\mu \nu}$ is expressed as 
\begin{equation}
	 \label{DepDeltaIm} \delta^{\mu \nu} = \sum_{i=1}^4 (Z^{-1})_{ji} \, P_i^\mu P_j^\nu, 
\end{equation}
where $Z^{-1}$ is the inverse of the Gram matrix $Z=[P_i \cdot P_j]_{i,j=1}^d $ and  $P_j^\mu\in\{p_1^\mu,p_2^\mu,p_3^\mu,n^\mu\}$.\\
Specifically, the Gram matrix takes the form
\begin{align}
Z=\left(\begin{matrix}
p_1^2&p_1\cdot p_2&p_1\cdot p_3&0\\
p_1\cdot p_2&p_2^2&p_2\cdot p_3&0\\
p_1\cdot p_3&p_2\cdot p_3&p_3^2&0\\
0&0&0&n^2
\end{matrix}\right),
\end{align}
where 
\begin{equation} 
	\label{DepDelta} 
	\delta^{\mu \nu} = \sum_{i,j=1}^3 \Big[(p_{i+1} \cdot p_{j-1} ) (p_{i-1} \cdot p_{j+1} ) - (p_{i-1} \cdot p_{j-1} ) (p_{i+1} \cdot p_{j+1} ) \Big] \frac{p_i^\mu p_j^\nu}{n^2} + \frac{n^\mu n^\nu}{n^2} , 
\end{equation}
with the indices labelled  mod-3. We show in Table \ref{4D} the number of tensor structures and independent form factors in $d = 4$ with the $n$-$p$ basis.

	\begin{table}[t]
	\centering
	\begin{tabular}{|c|c|c|} \hline&&\\[-2ex]
		Sector & $\#$ of tensor structures &$\#$ of orbits \\[1ex] \hline &&\\[-2ex]
		$n n n n p p$ & 4 & 2 \\[1ex]
		$n n p p p p$ &  73& 21 \\[1ex]
		$p p p p p p$ &  36 & 11 \\[1ex]\hline
			Total &113 &34 \\ \hline
	\end{tabular}
	\caption{ Number of tensor structures and independent form factors in $d=4$ with the $n$-$p$ basis. \label{4D}}
\end{table}
\section{Renormalizing the hierarchy}
In this section we are going to illustrate how the renormalization of the hierarchy works. 
The transverse traceless sector manifests a divergence at $d=4$ of the form $1/\epsilon$, with $\epsilon=d-4$. We expect that the form factors multiplying the tensorial structures with $(2\delta,2p)$ and $(3\delta)$ are divergent. We actually find, from the perturbative calculations, that the divergent part of the transverse traceless component is written, after some manipulation, as
	\begin{align}
	&\braket{t^{\mu_1\nu_1}t^{\mu_2\nu_2}j^{\mu_3}j^{\mu_4}}^{div}=\,\Pi^{(4-\varepsilon)\,\mu_1\nu_1}_{\alpha_1\beta_1}\Pi^{(4-\varepsilon)\,\mu_2\nu_2}_{\alpha_2\beta_2}\pi^{\mu_3}_{\alpha_3}\pi^{\mu_4}_{\alpha_4}\ \left(\frac{e^2\,(4N_f+N_s)}{12\pi^2\varepsilon}\right)\,\bigg\{  \delta^{\alpha_1 \alpha_2}\delta^{\beta_1 \beta_2} p_3^{\alpha_4}p_4^{\alpha_3}\notag\\[1ex]
	&-4\,\delta^{\alpha_1 \alpha_2} \delta^{\alpha_3 (\beta_2}p_4^{\beta_1)}p_3^{\alpha_4} +4 \delta^{\alpha_1 \alpha_2} \delta^{\alpha_3 \alpha_4}p_3^{(\beta_1} p_4^{\beta_2)}-4 \delta^{\alpha_1 \alpha_2} \delta^{\alpha_4 (\beta_2}p_3^{\beta_1)} p_4^{\alpha_3} 
	+2 \delta^{\alpha_2 \alpha_3} \delta^{\alpha_4 \beta_2}p_4^{\beta_1} p_3^{\alpha_1}+2\delta^{\alpha_1 \alpha_3} \delta^{\alpha_4 \beta_1}p_4^{\beta_2} p_3^{\alpha_2}\notag\\[1ex]
	&\quad-2 p_4^{\beta_1}p_3^{\beta_2} \delta^{\alpha_1 \alpha_3} \delta^{\alpha_2 \alpha_4}-2 p_4^{\beta_2} p_3^{\beta_1} \delta^{\alpha_1 \alpha_4} \delta^{\alpha_2 \alpha_3}+(s-p_3^2+p_4^2)\left[2\delta^{\alpha_1(\alpha_4}\delta^{\alpha_3)\alpha_2}\delta^{\beta_1\beta_2}-\frac{1}{2}\delta^{\alpha_1\alpha_2}\delta^{\beta_1\beta_2}\delta^{\alpha_3\alpha_4}\right]\bigg\},\label{ttjjDiv}
	\end{align}
where $N_f$ and $N_s$ indicate the the number of fermion and scalar families respectively, that are arbitrary. 
The projectors $\Pi$ are expanded around $d=4$ as
\begin{align}
\Pi^{(4-\varepsilon)\,\mu_1\nu_1}_{\alpha_1\beta_1}=\Pi^{(4)\,\mu_1\nu_1}_{\alpha_1\beta_1}-\frac{\varepsilon}{9}\,\pi^{\mu_1\nu_1}\pi_{\alpha_1\beta_1}+O(\varepsilon^2)
\end{align} 
with $\Pi^{(4)}$ the transverse traceless projectors defined in \eqref{TTproj} with $d=4$.

The counterterm vertex is 
\begin{equation}
\braket{T^{\mu_1\nu_1}T^{\mu_2\nu_2}J^{\mu_3}J^{\mu_4}}_{count}=-\frac{1}{\varepsilon}\sum_{I=f,s}\,N_I\,\beta_c(I)\,V_{F^2}^{\mu_1\nu_1\mu_2\nu_2\mu_3\mu_4}(p_1,p_2,p_3,p_4),
\end{equation}
where
\begin{align}
&V_{F^2}^{\mu_1\nu_1\mu_2\nu_2\mu_3\mu_4}(p_1,p_2,p_3,p_4)=4\int\,d^dx\prod_{k=1}^4d^dx_k\left(\frac{\delta^4\left(\sqrt{-g}\,F^2\right)(x)}{\delta g_{\mu_1\nu_1}(x_1)\delta g_{\mu_2\nu_2}(x_2)\delta A_{\mu_3}(x_3)\delta A_{\mu_4}(x_4)}\right)_{g\to\delta}\,e^{i\sum_j^4p_jx_j}
\end{align}
is the fourth functionla derivatives of the Lagrangian density $1/4 \sqrt{g} F^2$.
From this counterterm we can extract its transverse traceless part that can be written as
\begin{align}
	&\braket{t^{\mu_1\nu_1}t^{\mu_2\nu_2}j^{\mu_3}j^{\mu_4}}_{count}=\,\Pi^{(4-\varepsilon)\,\mu_1\nu_1}_{\alpha_1\beta_1}\Pi^{(4-\varepsilon)\,\mu_2\nu_2}_{\alpha_2\beta_2}\pi^{\mu_3}_{\alpha_3}\pi^{\mu_4}_{\alpha_4}\ \left(-\sum_{I=s,f}\frac{8}{\varepsilon}\,\beta_c(I)\,N_I\right)\,\bigg\{  \delta^{\alpha_1 \alpha_2}\delta^{\beta_1 \beta_2} p_3^{\alpha_4}p_4^{\alpha_3} \notag\\[1ex]
	&-4\,\delta^{\alpha_1 \alpha_2} \delta^{\alpha_3 (\beta_2}p_4^{\beta_1)}p_3^{\alpha_4} +4 \delta^{\alpha_1 \alpha_2} \delta^{\alpha_3 \alpha_4}p_3^{(\beta_1} p_4^{\beta_2)}-4 \delta^{\alpha_1 \alpha_2} \delta^{\alpha_4 (\beta_2}p_3^{\beta_1)} p_4^{\alpha_3} 
	+2 \delta^{\alpha_2 \alpha_3} \delta^{\alpha_4 \beta_2}p_4^{\beta_1} p_3^{\alpha_1}+2\delta^{\alpha_1 \alpha_3} \delta^{\alpha_4 \beta_1}p_4^{\beta_2} p_3^{\alpha_2}\notag\\[1ex]
	&\hspace{1.5cm}-2 p_4^{\beta_1}p_3^{\beta_2} \delta^{\alpha_1 \alpha_3} \delta^{\alpha_2 \alpha_4}-2 p_4^{\beta_2} p_3^{\beta_1} \delta^{\alpha_1 \alpha_4} \delta^{\alpha_2 \alpha_3}+(s-p_3^2+p_4^2)\left[2\delta^{\alpha_1(\alpha_4}\delta^{\alpha_3)\alpha_2}\delta^{\beta_1\beta_2}-\frac{1}{2}\delta^{\alpha_1\alpha_2}\delta^{\beta_1\beta_2}\delta^{\alpha_3\alpha_4}\right]\bigg\},\label{ttjjCount}
\end{align}
where  $s=(p_1+p_2)^2$. The divergences are removed with the choices
\begin{equation}
\beta_c(scalar)=\frac{e^2}{96\,\pi^2},\qquad \beta_c(fermion)=\frac{e^2}{24\,\pi^2}.
\end{equation}
It is worth mentioning, as expected, that these are exactly the same choices that renormalize the $2$-point function $\braket{JJ}$ and $3$-point function $\braket{TTJ}$ as well as all the other $n$-point functions involving two conserved currents.  
\subsection{Hierarchy of the counterterm contribution} 
Furthermore, the counterterm contribution satisfies a hierarchy of the form 
\begin{align}
	&p_{1\mu_1}\braket{T^{\mu_1\nu_1}(p_1)T^{\mu_2\nu_2}(p_2)J^{\mu_3}(p_3)J^{\mu_4}(p_4)}_{count}=\notag\\
	& 
	=\bigg[2\ {p_2}_{\lambda_1} \delta^{\nu_1 (\mu_2} \braket{T^{\nu_2) \lambda_1} (p_1+p_2) J^{\mu_3} (p_3) J^{\mu_4} (p_4)}_{count}  - 
	{p_2}^{\nu_1}  \braket{T^{\mu_2 \nu_2} (p_1+p_2) J^{\mu_3} (p_3) J^{\mu_4} (p_4)}_{count} \bigg] \notag\\ 
	&
	+ 2 \bigg\{  \bigg[ \delta^{\nu_1 (\mu_2} p_3^{\nu_2)}  \braket{J^{\mu_3} (p_1+p_2+p_3) J^{\mu_4} (p_4)}_{count} - \delta^{\nu_1 (\mu_2} \delta^{\nu_2) \mu_3}{p_3}_{\lambda_1 } \braket{J^{\lambda_1} (p_1+p_2+p_3) J^{\mu_4} (p_4)} _{count}
	\notag\\ 
	& + \, \frac{1}{2}\delta^{\mu_3 \nu_1} p_{3\lambda_1} \braket{J^{\lambda_1}(p_1+p_3) T^{\mu_2 \nu_2}(p_2) J^{\mu_4}(p_4) }_{count}- \frac{1}{2}p_3^{\nu_1 } \braket{J^{\mu_3}(p_1+p_3) T^{\mu_2 \nu_2}(p_2) J^{\mu_4}(p_4) }_{count} \bigg]+\bigg[ (3 \tor 4) \bigg]\bigg\} ,
\end{align}
generated by diffeomorphism invariance, 
where 
\begin{align}
\braket{J^{\mu_3}(p_3)J^{\mu_4}(p_4)}_{count}&=\left(-\sum_{I=s,f}\frac{1}{\varepsilon}\,\beta_c(I)\,N_I\right)\int d^dx\,d^dx_3\,d^dx_4\left(\frac{\delta^2\,F^2(x)}{\delta A_{\mu_3}(x_3)\delta A_{\mu_4}(x_4)}\right)\,e^{ip_3x_3+ip_4x_4}\notag\\
&=\left(-\sum_{I=s,f}\frac{4}{\varepsilon}\,\beta_c(I)\,N_I\right)\bigg[\delta^{\mu_3\mu_4}(p_3\cdot p_4)-p_4^{\mu_3}p_3^{\mu_4}\bigg]
\end{align}
is the counterterm 2-point function of two photons when $p_3=-p_4$. It is the counterterm of the photon self-energy at one-loop with intermediate scalars and fermions, as clear from the sum over 
$s$ and $f$ present in the equation above. 
The counterterm above renormalizes the $2$-point function $\braket{JJ}$, perturbatively expressed as
\begin{align}
\braket{J^{\mu_3}(-p_4)J^{\mu_4}(p_4)}=\frac{e^2}{(4\pi)^2}\frac{2(d-2)N_f+N_s}{(d-1)}\bigg[\delta^{\mu_3\mu_4}p_4^2-p_4^{\mu_3}p_4^{\mu_4}\bigg]\,B_0(p_4^2),
\end{align}
with $B_0(p_4^2)$ denoting the scalar (bubble) 2-point function, where the divergent part is extracted in DR as
\begin{align}
\braket{J^{\mu_3}(-p_4)J^{\mu_4}(p_4)}_{div}=\frac{e^2}{24\pi^2}\frac{4N_f+N_s}{\varepsilon}\bigg[\delta^{\mu_3\mu_4}p_4^2-p_4^{\mu_3}p_4^{\mu_4}\bigg].
\end{align}
Notice that the anomaly is generated by the trace of the counterterm due to the relation 

\begin{align}
	&\delta_{\mu_1\nu_1}\braket{T^{\mu_1\nu_1}(p_1)T^{\mu_2\nu_2}(p_2)J^{\mu_3}(p_3)J^{\mu_4}(p_4)}^{count}=\notag\\[1ex]
	&=2\big[\sqrt{g}\,F^2\big]^{\mu_2\nu_2\mu_3\mu_4}(p_2,p_3,p_4)-2 \, \braket{T^{\mu_2 \nu_2} (p_1+p_2) J^{\mu_3} (p_3) J^{\mu_4} (p_4) }_{count}.
\end{align}
The equations above clarifies the way the renormalization of the hierarchy occurs. By tracing the counterterm of the $TTJJ$ we generate an anomaly contribution, corresponding to the first term on the rhs, together with the counterterm of the 3-point function $TJJ$, recursively. 
Having identified the anomalous conservation and trace WIs satisfied by the correlator, the renormalization of the $TJJ$ in the hierarchy is ensured by the presence of the corresponding counterm, with its renormalized epxression given by
  
\begin{align}
\braket{ T^{\mu_1 \nu_1} (p_1)J^{\mu_3} (p_3) J^{\mu_4} (p_4) }_{Ren}&=\braket{ T^{\mu_1 \nu_1} (p_1)J^{\mu_3} (p_3) J^{\mu_4} (p_4) }_{fin}+\braket{ T^{\mu_1 \nu_1} (p_1)J^{\mu_3} (p_3) J^{\mu_4} (p_4) }_{anom}
\end{align}
where the first contribution on the rhs is the finite expression at $d=4$, while its anomaly contribution is given by 
\begin{align}
\braket{ T^{\mu_1 \nu_1} (p_1)J^{\mu_3} (p_3) J^{\mu_4} (p_4) }_{anom}=\sum_{I=s,f}\beta_c(I)\,\frac{\pi^{\mu_1\nu_1}(p_1)}{3}\Big[F^2\Big]^{\mu_3\mu_4}(p_3,p_4).\label{anomTJJ}
\end{align}
\subsection{The reconstruction of the renormalized $TTJJ$ and the trace anomaly }
The renormalized correlator is given by

\begin{align}
&\braket{ T^{\mu_1 \nu_1} (p_1) T^{\mu_2 \nu_2} (p_2) J^{\mu_3} (p_3) J^{\mu_4} (p_4) }_{Ren}=\notag\\ &=\Big(\braket{ T^{\mu_1 \nu_1} (p_1) T^{\mu_2 \nu_2} (p_2) J^{\mu_3} (p_3) J^{\mu_4} (p_4) }+\braket{ T^{\mu_1 \nu_1} (p_1) T^{\mu_2 \nu_2} (p_2) J^{\mu_3} (p_3) J^{\mu_4} (p_4) }_{count} \Big)_{d\to4}\notag\\
&=\braket{ T^{\mu_1 \nu_1} (p_1) T^{\mu_2 \nu_2} (p_2) J^{\mu_3} (p_3) J^{\mu_4} (p_4) }_{fin}+\braket{ T^{\mu_1 \nu_1} (p_1) T^{\mu_2 \nu_2} (p_2) J^{\mu_3} (p_3) J^{\mu_4} (p_4) }_{anom},
\end{align}
where the bare correlator and the counterterm are re-expressed in terms of a finite renormalized correlator not contributing to the trace Ward identity, and a second part which accounts for the trace anomaly.\\
Obviously, the renormalization procedure involves also an analysis of the local sector, as clear from the reconstruction given in \eqref{loc}. An example is

\begin{align}
&\braket{ t_{loc}^{\mu_1 \nu_1}T^{\mu_2 \nu_2} J^{\mu_3} J^{\mu_4}}_{Ren}=\bigg(\braket{ t_{loc}^{\mu_1 \nu_1}T^{\mu_2 \nu_2} J^{\mu_3} J^{\mu_4}}_{div}+\braket{ t_{loc}^{\mu_1 \nu_1}T^{\mu_2 \nu_2} J^{\mu_3} J^{\mu_4}}_{count}\bigg)_{d\to4}\notag\\
&=\braket{ t_{loc}^{\mu_1 \nu_1}T^{\mu_2 \nu_2} J^{\mu_3} J^{\mu_4}}_{fin}^{(d=4)}+\braket{ t_{loc}^{\mu_1 \nu_1}T^{\mu_2 \nu_2} J^{\mu_3} J^{\mu_4}}_{anom}^{(d=4)},
\end{align}

where 
\begin{align}
	&\braket{ t_{loc}^{\mu_1 \nu_1}T^{\mu_2 \nu_2} J^{\mu_3} J^{\mu_4}}_{anom}^{(d=4)}=\notag\\
&=\beta_C\,\bigg\{\mathcal{I}^{(d=4)\,\mu_1\nu_1}_{\alpha_1}\bigg[2\ {p_2}_{\lambda_1}  \frac{\delta^{\alpha_1 (\mu_2}\pi^{\nu_2) \lambda_1} (p_1+p_2)}{3}\left[F^2\right]^{\mu_3\mu_4}(p_3,p_4) - 
{p_2}^{\alpha_1}  \frac{\pi^{\mu_2 \nu_2} (p_1+p_2) }{3}\left[F^2\right]^{\mu_3\mu_4}(p_3,p_4) \notag\\ 
&
+ \frac{\pi^{\mu_2\nu_2}(p_2)}{3}\bigg( \delta^{\mu_3 \alpha_1} p_{3\lambda_1} \left[F^2\right]^{\lambda_1\mu_4}(p_1+p_3,p_4)- p_3^{\alpha_1 } \left[F^2\right]^{\mu_3\mu_4}(p_1+p_3,p_4)\bigg)\notag\\
&+ \frac{\pi^{\mu_2\nu_2}(p_2)}{3}\bigg( \delta^{\mu_4 \alpha_1} p_{4\lambda_1} \left[F^2\right]^{\lambda_1\mu_3}(p_1+p_4,p_3)- p_4^{\alpha_1 } \left[F^2\right]^{\mu_3\mu_4}(p_3,p_1+p_4)\bigg)\bigg]\notag\\
&-2\,\frac{\pi^{\mu_1\nu_1}(p_1)}{3}\frac{\pi^{\mu_2\nu_2}(p_1+p_2)}{3} \left[F^2\right]^{\mu_3\mu_4}(p_3,p_4)+2\frac{\pi^{\mu_1\nu_1}(p_1)}{3}\big[\sqrt{g}\,F^2\big]^{\mu_2\nu_2\mu_3\mu_4}(p_2,p_3,p_4)\bigg\}\label{tlocTJJ},
\end{align}
Combining all these expressions, we identify a renormalized hierarchy stemming from diffeomorphism invariance. This hierarchy splits into two distinct contributions: one adhering to an ordinary hierarchy and the other constituting the anomalous hierarchy, which consists of two parts. The first part of this anomalous hierarchy corresponds to a traceless correlator, referred to as the "$0$-residue term". The second part, termed the "anomaly pole contribution", is structured around Fourier transforms of $R\Box^{-1}$ operators, which in flat space are directly proportional to $\pi$ projectors. Both hierarchies satisfy diffeomorphism invariance. 
\subsection{The anomalous hierarchy}
Let's now focus on the anomalous hierarchy. The correlator that identifies the hierarchical equations is organised in the form 

\begin{align}
\braket{ T^{\mu_1 \nu_1}T^{\mu_2 \nu_2} J^{\mu_3} J^{\mu_4}}_{anom}^{(d=4)}&=\braket{ t_{loc}^{\mu_1 \nu_1}T^{\mu_2 \nu_2} J^{\mu_3} J^{\mu_4}}_{anom}^{(d=4)}+\braket{T^{\mu_1 \nu_1}t_{loc}^{\mu_2 \nu_2} J^{\mu_3} J^{\mu_4}}_{anom}^{(d=4)}-\braket{ t_{loc}^{\mu_1 \nu_1}t_{loc}^{\mu_2 \nu_2} J^{\mu_3} J^{\mu_4}}_{anom}^{(d=4)}\notag\\
&=\braket{T^{\mu_1 \nu_1}T^{\mu_2 \nu_2} J^{\mu_3} J^{\mu_4}}_{0-residue}+\braket{ T^{\mu_1 \nu_1}T^{\mu_2 \nu_2} J^{\mu_3} J^{\mu_4}}_{pole}\label{resultFinal}.
 \end{align}
which are the two contributions mentioned above. Notice that the "0-residue" part 

\begin{align}
	&\braket{T^{\mu_1 \nu_1}T^{\mu_2 \nu_2} J^{\mu_3} J^{\mu_4}}_{0-residue}=\notag\\
	&=\beta_C\,\bigg\{\mathcal{I}^{(d=4)\,\mu_1\nu_1}_{\alpha_1}\Pi^{\mu_2\nu_2}_{\alpha_2\beta_2}\left(\frac{2}{3}\ {p_2}_{\lambda_1}  \delta^{\alpha_1 \alpha_2}\pi^{\beta_2\lambda_1} (p_1+p_2)- 
	\frac{1}{3}\,{p_2}^{\alpha_1}  \pi^{\alpha_2 \beta_2} (p_1+p_2) \right)\left[F^2\right]^{\mu_3\mu_4}(p_3,p_4)\notag\\
	&\hspace{1.5cm}+\mathcal{I}^{(d=4)\,\mu_2\nu_2}_{\alpha_2}\Pi^{\mu_1\nu_1}_{\alpha_1\beta_1}\left(\frac{2}{3}\ {p_1}_{\lambda_2}  \,\delta^{\alpha_1 \alpha_2}\pi^{\beta_1\lambda_2} (p_1+p_2)- 
	\frac{1}{3}{p_1}^{\alpha_2}  \,\pi^{\alpha_1 \beta_1} (p_1+p_2) \right)\left[F^2\right]^{\mu_3\mu_4}(p_3,p_4)\notag\\	
	&\hspace{1.5cm}+\mathcal{I}^{(d=4)\,\mu_1\nu_1}_{\alpha_1}\mathcal{I}^{(d=4)\,\mu_2\nu_2}_{\alpha_2}\ \delta^{\alpha_1\alpha_2}\,p_{2\lambda_1}\,p_{2\lambda_2}\frac{\pi^{\lambda_1\lambda_2}(p_1+p_2)}{3}\,\Big[F^2\Big]^{\mu_3\mu_4}(p_3,p_4)\bigg\},\label{0-residue}
\end{align}
corresponds to a Weyl invariant contribution, for being traceless 
with the property
\begin{align}
&\delta_{\mu_i\nu_i}\braket{ T^{\mu_1 \nu_1}T^{\mu_2 \nu_2} J^{\mu_3} J^{\mu_4}}_{0-residue}=0,\\
\end{align}
but plays an essential role in order for the anomalous part of the correlator 
$\braket{ T^{\mu_1 \nu_1}T^{\mu_2 \nu_2} J^{\mu_3} J^{\mu_4}}_{anom}$ to respect the correct WI from diffeomorphism invariance. Indeed, such term is non homogenous and contributes to the longitudinal sector

\begin{align}
&p_{1\mu_1}\braket{ T^{\mu_1 \nu_1}T^{\mu_2 \nu_2} J^{\mu_3} J^{\mu_4}}_{0-residue}=\notag\\
&=\beta_C\,\bigg\{\Pi^{\mu_2\nu_2}_{\alpha_2\beta_2}\left(\frac{2}{3}\ {p_2}_{\lambda_1}  \delta^{\nu_1 \alpha_2}\pi^{\beta_2\lambda_1} (p_1+p_2)- 
\frac{1}{3}\,{p_2}^{\nu_1}  \pi^{\alpha_2 \beta_2} (p_1+p_2) \right)\left[F^2\right]^{\mu_3\mu_4}(p_3,p_4)\notag\\
&\hspace{1.5cm}+\ \delta^{\nu_1\alpha_2}\,p_{2\lambda_1}\,p_{1\lambda_2}\frac{\pi^{\lambda_1\lambda_2}(p_1+p_2)}{3}\,\Big[F^2\Big]^{\mu_3\mu_4}(p_3,p_4)\bigg\},\notag\\
&p_{1\mu_1}p_{2\mu_2}\braket{ T^{\mu_1 \nu_1}T^{\mu_2 \nu_2} J^{\mu_3} J^{\mu_4}}_{0-residue}=\beta_C\,\bigg\{\delta^{\nu_1\alpha_2}\,p_{2\lambda_1}\,p_{2\lambda_2}\frac{\pi^{\lambda_1\lambda_2}(p_1+p_2)}{3}\,\Big[F^2\Big]^{\mu_3\mu_4}(p_3,p_4)\bigg\}.\label{0residue}
\end{align}
The renormlization of 3-point functions has so far shown, in all the previous analysis mentioned above that  
the anomalous WIs are directly linked to "anomaly poles", extracted from the longitudinal projector $\pi$, in the form 
\beq
\pi^{\mu\nu}(p)=\frac{1}{p^2}\hat{\pi}^{\mu\nu}\qquad\qquad  \hat{\pi}^{\mu\nu}=p^2 g^{\mu\nu}-p^\mu p^\nu
 \eeq
 with $1/p^2$, the pole, turning into a $\Box^{-1}$ term in coordinate space. 
One can verify by a lengthy evaluation that the pole part that
\begin{align}
\braket{ T^{\mu_1 \nu_1}T^{\mu_2 \nu_2} J^{\mu_3} J^{\mu_4}}_{pole}&=\beta_C\ \bigg\{2\frac{\pi^{\mu_1\nu_1}(p_1)}{3}\left(\big[\sqrt{g}\,F^2\big]^{\mu_2\nu_2\mu_3\mu_4}(p_2,p_3,p_4)-\frac{\pi^{\mu_2\nu_2}(p_1+p_2)}{3} \left[F^2\right]^{\mu_3\mu_4}(p_3,p_4)\right)\notag\\
&+2\frac{\pi^{\mu_2\nu_2}(p_2)}{3}\left(\big[\sqrt{g}\,F^2\big]^{\mu_1\nu_1\mu_3\mu_4}(p_1,p_3,p_4)-\frac{\pi^{\mu_1\nu_1}(p_1+p_2)}{3} \left[F^2\right]^{\mu_3\mu_4}(p_3,p_4)\right)\notag\\
&+2\,\frac{\pi^{\mu_1\nu_1}(p_1)}{3}\frac{\pi^{\mu_2\nu_2}(p_2)}{3}\,\Big[F^2\Big]^{\mu_3\mu_4}(p_3,p_4)\bigg\}\label{pole}
\end{align}
satisfies the correct expression of the trace hierarchy

\begin{align}
\delta_{\mu_i\nu_i}\braket{ T^{\mu_1 \nu_1}T^{\mu_2 \nu_2} J^{\mu_3} J^{\mu_4}}_{pole}&=2\beta_C\,\left(\big[\sqrt{g}\,F^2\big]^{\mu_j\nu_j\mu_3\mu_4}(p_j,p_3,p_4)-\frac{\pi^{\mu_j\nu_j}(p_1+p_2)}{3} \left[F^2\right]^{\mu_3\mu_4}(p_3,p_4)\right),
\end{align}
but not the conservation WI hierarchy, generated by requiring diffeomorphism invariance, since  
\begin{align}
p_{\mu_i}\braket{ T^{\mu_1 \nu_1}T^{\mu_2 \nu_2} J^{\mu_3} J^{\mu_4}}_{pole}&=\frac{2\beta_C\,\pi^{\mu_j\nu_j}(p_j)}{3}\,\,p_{i\mu_i}\left(\big[\sqrt{g}\,F^2\big]^{\mu_i\nu_i\mu_3\mu_4}(p_i,p_3,p_4)\right.\notag\\
&\left. -\frac{\pi^{\mu_i\nu_i}(p_1+p_2)}{3}\left[F^2\right]^{\mu_3\mu_4}(p_3,p_4)\right),
\end{align}
differs from the correct expression of the hierarchy that is expected to be satisfied by the anomalous sector
\begin{align}
\label{int}
	&p_{1\mu_1}\braket{T^{\mu_1\nu_1}(p_1)T^{\mu_2\nu_2}(p_2)J^{\mu_3}(p_3)J^{\mu_4}(p_4)}_{anom}=\notag\\
	& 
	=\bigg[2\ {p_2}_{\lambda_1} \delta^{\nu_1 (\mu_2} \braket{T^{\nu_2) \lambda_1} (p_1+p_2) J^{\mu_3} (p_3) J^{\mu_4} (p_4)}_{anom}  - 
	{p_2}^{\nu_1}  \braket{T^{\mu_2 \nu_2} (p_1+p_2) J^{\mu_3} (p_3) J^{\mu_4} (p_4)}_{anom} \bigg] \notag\\ 
	&
	+  \bigg\{  \bigg[  \, \delta^{\mu_3 \nu_1} p_{3\lambda_1} \braket{J^{\lambda_1}(p_1+p_3) T^{\mu_2 \nu_2}(p_2) J^{\mu_4}(p_4) }_{anom}- p_3^{\nu_1 } \braket{J^{\mu_3}(p_1+p_3) T^{\mu_2 \nu_2}(p_2) J^{\mu_4}(p_4) }_{anom} \bigg]	+\bigg[ (3 \tor 4) \bigg]\bigg\}.
\end{align}
The perturbative analysis of the conservation WI provides the template that the AIA should respect, since 
$\braket{T^{\mu_1\nu_1}(p_1)T^{\mu_2\nu_2}(p_2)J^{\mu_3}(p_3)J^{\mu_4}(p_4)}_{anom}$ contains naturally, in the perturbative diagrammatic expansion both the $0-residue$ term and the pole part. 

\subsection{The anomaly-pole structure of the $TTT$}
\label{anompole}
\begin{figure}[t]
\centering
\subfigure[]{\includegraphics[scale=0.49]{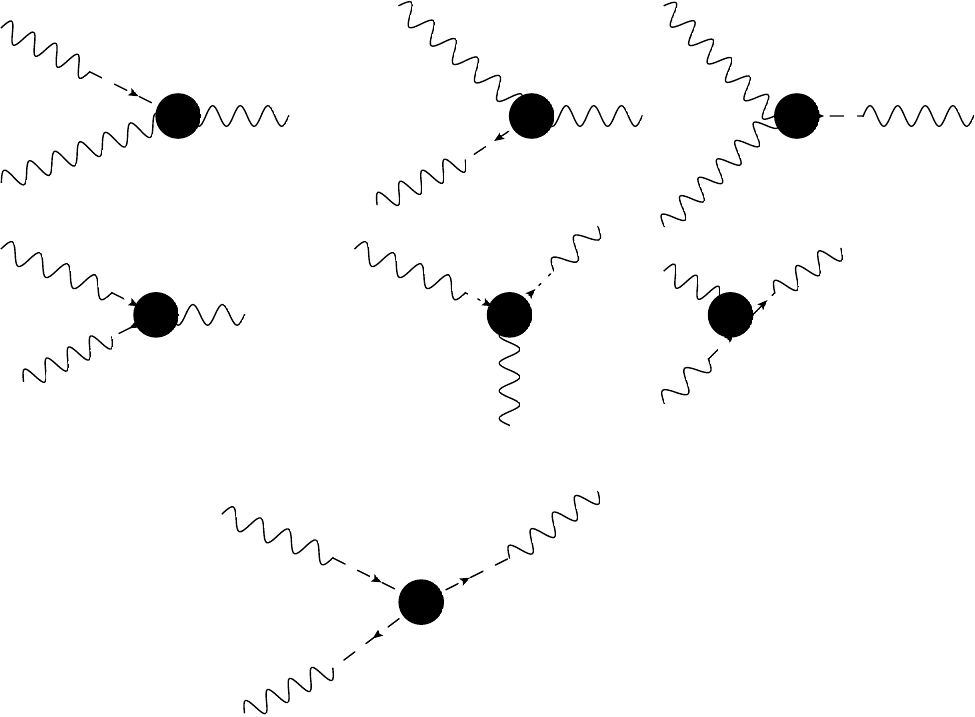}} 
\caption{Anomaly interactions mediated by the exchange of one, two or three poles in the 3-graviton vertex.  They account for $TTT_{anom}$ and there are no Weyl-invariant terms to be included. The same vertex is predicted by the AIAs.}
\label{dec}
\end{figure}

We show in Fig. \eqref{dec} a pictorial description of the structure of the anomalous part of the $TTT$. In this case, as already mentioned in the Introduction, there are no missing Weyl-invariant terms in the hierarchy of this 3-point function. The anomalous hierarchies of this correlator, either built around diffeomorphism invariance, special conformal invariance or dilatation invariance, are consistently defined and are described 
pictorially in this figure. The dashed external lines in the figure symbolize the insertion of external $\pi^{\mu\nu}$ projectors, from which we have extracted an anomaly pole $(1/p^2)$. These contributions, in coordinate space, indicate insertions of a $R^{(1)}\Box^{-1}$ bilinear vertex on the external lines, coupled to derivatives of the anomaly functional, with $R^{(1)}$ denoting   
the linearized scalar curvature. By tracing the indices of each of these transverse projectors we obtain a trace contribution. The anomaly functional is therefore reinterpreted as the residue at the $1/p^2$ pole once we trace the indices of the projector. A "0-residue" term, as we have defined it, is therefore deprived of such external projectors. The projectors are linked to trilinear vertices contructed by functional derivatives of the anomaly functional.  We have

\begin{align}
\braket{T(p_1)T^{\mu_2\nu_2}(p_2)T^{\mu_3\nu_3}(\bar{p}_3)}_{anom}^{(4)}&=\big(4[E]^{\m_2\n_2\m_3\n_3}(p_2,p_3)+4[C^2]^{\m_2\n_2\m_3\n_3}(p_2,\bar{p}_3)\big)\\
\braket{T(p_1)T(p_2)T^{\mu_3\nu_3}(\bar{p}_3)}_{anom}^{(4)}&=\d_{\a_2\b_2}\big(4[E]^{\a_2\b_2\m_3\n_3}(p_2,p_3)+4[C^2]^{\a_2\b_2\m_3\n_3}(p_2,\bar{p}_3)\big)\notag\\
\braket{T(p_1)T(p_2)T(\bar{p}_3)}_{an}^{(4)}&=\d_{\a_2\b_2}\d_{\a_3\b_3}\big(4[E]^{\a_2\b_2\a_3\b_3}(p_2,p_3)+4[C^2]^{\a_2\b_2\a_3\b_3}(p_2,\bar{p}_3)\big)
\end{align}   
where we have denoted by $[E]^{\m_2\n_2\m_3\n_3}$ and $[C^2]^{\a_2\b_2\m_3\n_3}$ the second functional derivatives of the Gauss Bonnet tensor $E$ and of the Weyl tensor squared $C^2$.
The anomaly part, in the case of the $TTT$, is only given by the pole terms 
\begin{align}
\label{expans}
& \lla T^{\mu_1 \nu_1}({p}_1) T^{\mu_2 \nu_2}({p}_2) T^{\mu_3 \nu_3}({p}_3) \rra_{an} = \frac{\hat\pi^{\mu_1 \nu_1}({p}_1)}{3\, p_1^2} \lla T({p}_1) T^{\mu_2 \nu_2}({p}_2) T^{\mu_3 \nu_3}({p}_3) \rra_{an} \nn
& + \frac{\hat\pi^{\mu_2 \nu_2}({p}_2)}{3\, p_2^2} \lla T^{\mu_1 \nu_1}({p}_1) T({p}_2) T^{\mu_3 \nu_3}({p}_3) \rra_{an} + \frac{\hat\pi^{\mu_3 \nu_3}({p}_3)}{3\, p_3^2} \lla T^{\mu_1 \nu_1}({p}_1) T^{\mu_2 \nu_2}({p}_2)  T({p}_3)\rra_{an} \notag \\
& - \: \frac{\hat\pi^{\mu_1 \nu_1}({p}_1) \hat\pi^{\mu_2 \nu_2}({p}_2)}{9\, p_1^2 p_2^2}\lla T({p}_1)T({p}_2)T^{\mu_3 \nu_3}({p}_3)\rra_{an} - \: \frac{\hat\pi^{\mu_2 \nu_2}({p}_2) \hat\pi^{\mu_3 \nu_3}({p}_2)}{9 p_2^2 p_3^2}\lla T^{\mu_1 \nu_1}({p}_1)T(p_2)T({p}_3)\rra_{an} \notag \\
& - \: \frac{\hat\pi^{\mu_1 \nu_1}({p}_1)\hat \pi^{\mu_3 \nu_3}(\bar{p}_3)}{9 p_1^2 p_3^2}\lla T({p}_1) T^{\mu_2 \nu_2}({p}_2)T({p}_3)\rra_{an}  + \frac{\hat\pi^{\mu_1 \nu_1}({p}_1)\hat\pi^{\mu_2 \nu_2}({p}_2) \hat\pi^{\mu_3 \nu_3}(\bar{p}_3)}{27 p_1^2 p_2^2 p_3^2}\lla T({p}_1)T({p}_2)T(\bar{p}_3)\rra_{an} .
\end{align}
The consistency between this anomaly part of the $TTT$ correlator and the anomalous part of the hierarchy obtained by the corresponding pertturbative analysis of the $TTT$ has been shown in previous works 
\cite{Coriano:2017mux}.
 
\subsection{AIAs} 
Now, we move our focus towards an examination of the $TTJJ$ correlator computed by the functional differentiation of anomaly-induced actions. \\
We represent the functional denoted by $\mathcal{S}_A[g]$ as the integration of the trace anomaly
\begin{equation}
2\,g_{\mu\nu}\,\frac{\delta\,\,\mathcal{S}_A[g]}{\delta g_{\mu\nu}(x)}=\sqrt{-g}\bigg[b\,C^2+b'\,E+\beta_C\,F^2\bigg].
\end{equation}
This functional arises from a Weyl rescaling of the metric, $g_{\mu\nu}(x)=e^{2\phi(x)}\,\bar{g}_{\mu\nu}(x)$, leading to
\begin{align}
	\mathcal{S}_A[g]=\mathcal{S}[\bar g]+\Delta\,\mathcal{S}[\sigma,\bar g],
\end{align}
such that its conformal variation
\begin{equation}
2\,g_{\mu\nu}\,\frac{\delta\,\mathcal{S}_A[g]}{\delta g_{\mu\nu}(x)}=\frac{\delta\,\mathcal{S}_A[g]}{\delta \phi(x)}\Bigg|_{g=e^{2\phi}\bar{g}}=\frac{\delta\,\big(\Delta \mathcal{S}[\phi,\bar g]\big)}{\delta \phi(x)}=\sqrt{-g}\bigg[b\,C^2+b'\,E+\beta_C\,F^2\bigg]\Bigg|_{g=e^{2\phi}\bar{g}}\label{Weylconstr}
\end{equation}
represents the anomaly. The approach is well-known and has been reviewed in \cite{Coriano:2020ees}. After integrating \eqref{Weylconstr} along a path $\tilde{\phi}(x,\lambda)=\lambda\,\phi(x)$ with $0\ge\lambda\ge1$, we obtain the explicit expression for $\Delta \mathcal{S}$ as
\begin{align}
\Delta\mathcal{S}[\phi,\bar g]&=\int\,d^4x\int_0^1\,d\lambda\,\frac{\delta\,\big(\Delta \mathcal{S}[\phi,\bar g]\big)}{\delta \phi(x)}\Bigg|_{\phi=\tilde{\phi}(x,\lambda)}\frac{\partial\,\tilde{\phi}(x,\lambda)}{\partial\lambda}\notag\\
&=\int\,d^4x\int_0^1\,d\lambda\,\bigg[\sqrt{-g}\bigg(b\,C^2+b'\,E+\beta_C\,F^2\bigg)\bigg]_{g=e^{2\tilde\phi(x,\lambda)}\bar{g}}\,\phi(x).
\end{align}
Following the integration, $\Delta \mathcal{S}$ can be expressed as
\begin{align}
	\Delta \mathcal{S}[\phi,\bar g]&=b'\,\int\,d^4x\,\sqrt{-\bar g}\bigg[2\,\phi\,\bar\Delta_4\phi+\bigg(\bar E-\frac{2}{3}\bar \square\bar R\bigg)\phi\bigg]+\int d^4x\,\sqrt{-\bar g}\bigg[b\,\bar C^2+\beta_C\,\bar F^2\bigg]\phi,\label{deltaW}
\end{align}
except for terms independent of $\sigma$, i.e., conformally invariant terms, which do not contribute to the variation \eqref{Weylconstr}. In \eqref{deltaW}, we introduce the fourth-order operator $\Delta_4$ as
\begin{equation}
	\Delta_4=\nabla_{\mu}\left(\nabla^\mu\nabla^\nu+2R^{\mu\nu}-\frac{2}{3}R\,g^{\mu\nu}\right)\nabla_\nu=\square^2+2R^{\mu\nu}\,\nabla_{\mu}\nabla_\nu-\frac{2}{3}R\,\square+\frac{1}{3}\,\nabla^\mu R\,\nabla_{\mu}.\label{Delta4}
\end{equation}
To derive a nonlocal form of the anomaly-induced action, one can impose a condition $\chi(g)=0$, ensuring the satisfaction of $\chi(g\,e^{-2\phi})=0$. From this condition, we solve for $\phi$ in terms of a function of the metric, mainly $\phi=\Sigma(g)$, yielding
\begin{equation}
	\bar{g}_{\mu\nu}=e^{-2\,\Sigma(g)}\,g_{\mu\nu}, \quad g_{\alpha\beta}\frac{\delta}{\delta g_{\alpha\beta}}\,\bar{g}_{\mu\nu}[g]=0.
\end{equation}
Thus, the conformal decomposition is expressed as 
\begin{align}
	\mathcal{S}[g]&=\bar{\mathcal{S}}[g]+\mathcal{S}_A[g,\Sigma(g)]\\
	\mathcal{S}_A[g,\Sigma(g)]&=\,\int\,d^4x\,\sqrt{- g}\bigg[b'\bigg( E-\frac{2}{3} \square R\bigg)\Sigma+\bigg(b\, C^2+\beta_C\, F^2\bigg)\Sigma-2 b'\,\Sigma\,\Delta_4\Sigma\bigg]\label{WA}
\end{align}
where $\bar{\mathcal{S}}[g]$ and $\mathcal{S}_A[g,\Sigma(g)]$ represent its conformal-invariant and anomalous parts, respectively.
There exist two distinct exactly solvable conformal choices in $4D$, in which the gauge parameter $\Sigma(g)$ can be calculated in a closed form as a functional of $g$. They are discussed in \cite{Barvinsky:1995it}. One arises from Fradkin and Vilkovisky (FV) \cite{Fradkin:1978yw}
\begin{align}
	\chi_{FV}(g)=R(g)
\end{align}
with 
\begin{align}
	\Sigma_{FV}=-\ln\left(1+\frac{1}{6}(\square-R/6)^{-1}\,R\right)\label{FVgauge}
\end{align}
where $(\square-R/6)^{-1}$ denotes the inverse of the operator $\square-R/6$, playing the role of the corresponding Faddeev-Popov operator. Another choice arises from Riegert \cite{Riegert:1984kt} with
\begin{align}
	\chi_{R}(g)=E(g)-\frac{2}{3}\square\,R(g),
\end{align}
for which
\begin{align}
	\Sigma_{R}=\frac{1}{4}\,(\Delta_4)^{-1}\left(E-\frac{2}{3}\square\,R\right),\label{Rgauge}
\end{align}
where $(\Delta_4)^{-1}$ represents the inverse of the operator in \eqref{Delta4}. A direct analysis of the $TTJJ$ shows that the hierarchy derived in the Riegert gauge does not admit an ordinary Fourer transform to momentum space without the inclusion of a regularization cutoff. This is due to the appearance of double poles, while in the Fradkin-Vilkovisky gauge the correlator does not satisfy the correct trace Ward identities. \\
It can be shown, though, that the correct expression of the anomalous hierarchy, for the $TTJJ$ correlator, can be obtained by varying the following functional
\begin{align}
	\mathcal{S}_{anom}^{(2)}&=-\frac{\beta_C}{6} \int d^4 x \int d^4 x' \, \bigg\{\,
	 (\sqrt{-g} \, F^2)_x^{(1)} \left( {1 \over \square_0} \right)_{xx'}R^{(1)}_{x'}  
	+ \, F^2_x \left( {1 \over \square_0} \right)_{xx'} R^{(2)}_{x'} 
	\notag\\ &\hspace{-2cm}
	+\int d^4 x'' \bigg[\, F^2_x \left( {1 \over \square_0} \right)_{xx'}( \square_1)_{x'} \left( {1 \over \square_0} \right)_{x'x''}R^{(1)}_{x''} -\frac{1}{6}\, F^2_x \left( {1 \over \square_0} \right)_{xx'}\,R^{(1)}_{x'}\, \left( {1 \over \square_0} \right)_{x'x''}R^{(1)}_{x''}  \notag\\
	&+\frac{1}{3}R^{(1)}_x\left( {1 \over \square_0} \right)_{xx'}\,F^2_{x'}\left( {1 \over \square_0} \right)_{x'x''}R^{(1)}_{x''}\bigg]\bigg\},
\end{align}
which has been identified by resorting to the perturbative analysis in flat space. 

\section{Conclusions} 
Conformal back-reaction stands as an important possibility in the framework of cosmological models, wherein the trace of the stress-energy tensor driving the cosmological evolution is attributed not to the conventional cosmological constant but rather to the trace anomaly, characterising  the breaking of Weyl invariance in the early universe. This analytical paradigm offers the possibility of studying dark energy in a dynamical context. 
This perspective finds its expression in nonlocal actions, commonly termed AIAs. \\
The expansion of these nonlocal actions around Minkowski space is anticipated to yield semiclassical correlators subject to hierarchical Ward identities, intricately linked to both the anomalous conformal symmetry and diffeomorphism invariance.\\
In our inquiry, we direct our attention to the hierarchical structures inherent in a specific 4-point function, namely the 2-graviton-2-photon correlator (TTJJ), within the domain of free field theory. Our focus lies particularly on unraveling the structural intricacies of its hierarchy stemming from diffeomorphism invariance.\\
This hierarchy naturally partitions into an ordinary, non-anomalous component and an anomalous counterpart. Leveraging recent approaches from CFT in momentum space, we extend these techniques to the domain of 4-point functions, conducting a general classification of the tensorial structures and form factors residing within its transverse traceless sector. We demonstrate the split of the TTJJ correlator's hierarchy into both an ordinary and an anomalous part, both of which respect the conservation Ward Identities stemming from diffeomorphism invariance. However, our analysis unveils that the anomalous hierarchy does not reproduce the structure 
predicted by the perturbative analysis. The free field theory approach shows that starting from 4-point functions, some Weyl-invariant terms contributions are essential for ensuring its consistency. \\
It is still open the problem of how to render these AIAs consistent on general grounds. 
Since the attention of the theory community on nonlocal variants of GR is significant \cite{Capozziello:2024qol,Capozziello:2021krv}\cite{Maggiore:2014sia}, AIAs may provide an important link in order to relate nonlocal cosmologies to a fundamental symmetry such as conformal symmety.

\centerline{\bf Acknowledgements}
This work is partially supported by INFN within the Iniziativa Specifica QG-sky.  
The work of C. C. is funded by the European Union, Next Generation EU, PNRR project "National Centre for HPC, Big Data and Quantum Computing", project code CN00000013. This work is partially supported by the the grant PRIN 2022BP52A MUR "The Holographic Universe for all Lambdas" Lecce-Naples.

\providecommand{\href}[2]{#2}\begingroup\raggedright\endgroup


\begin{thebibliography}{10}

\bibitem{Antoniadis:2006wq}
I.~Antoniadis, P.~O. Mazur, and E.~Mottola, {\it {Cosmological dark energy:
  Prospects for a dynamical theory}},  {\em New J. Phys.} {\bf 9} (2007) 11,
  [\href{http://xxx.lanl.gov/abs/gr-qc/0612068}{{\tt gr-qc/0612068}}].

\bibitem{Lucat:2018slu}
S.~Lucat, T.~Prokopec, and B.~Swiezewska, {\it {Conformal symmetry and the
  cosmological constant problem}},  {\em Int. J. Mod. Phys. D} {\bf 27} (2018),
  no.~14 1847014, [\href{http://xxx.lanl.gov/abs/1804.00926}{{\tt
  arXiv:1804.00926}}].

\bibitem{Pelinson:2010yr}
A.~M. Pelinson and I.~L. Shapiro, {\it {On the scaling rules for the
  anomaly-induced effective action of metric and electromagnetic field}},  {\em
  Phys. Lett. B} {\bf 694} (2011) 467--472,
  [\href{http://xxx.lanl.gov/abs/1005.1313}{{\tt arXiv:1005.1313}}].

\bibitem{Ghosh:2020qsx}
J.~K. Ghosh, E.~Kiritsis, F.~Nitti, and L.~T. Witkowski, {\it {Back-reaction in
  massless de Sitter QFTs: holography, gravitational DBI action and f(R)
  gravity}},  {\em JCAP} {\bf 07} (2020) 040,
  [\href{http://xxx.lanl.gov/abs/2003.09435}{{\tt arXiv:2003.09435}}].

\bibitem{Coriano:2021nvn}
C.~Corian\`o, M.~M. Maglio, and D.~Theofilopoulos, {\it {The conformal anomaly
  action to fourth order (4T) in $d=4$ in momentum space}},  {\em Eur. Phys. J.
  C} {\bf 81} (2021), no.~8 740,
  [\href{http://xxx.lanl.gov/abs/2103.13957}{{\tt arXiv:2103.13957}}].

\bibitem{Coriano:2018bsy}
C.~Corian\`o and M.~M. Maglio, {\it {The general 3-graviton vertex ($TTT$) of
  conformal field theories in momentum space in $d =4$}},  {\em Nucl. Phys.}
  {\bf B937} (2018) 56--134, [\href{http://xxx.lanl.gov/abs/1808.10221}{{\tt
  arXiv:1808.10221}}].

\bibitem{Coriano:2017mux}
C.~Corian\`o, M.~M. Maglio, and E.~Mottola, {\it {TTT in CFT: Trace Identities
  and the Conformal Anomaly Effective Action}},  {\em Nucl. Phys.} {\bf B942}
  (2019) 303--328, [\href{http://xxx.lanl.gov/abs/1703.08860}{{\tt
  arXiv:1703.08860}}].

\bibitem{Bzowski:2013sza}
A.~Bzowski, P.~McFadden, and K.~Skenderis, {\it {Implications of conformal
  invariance in momentum space}},  {\em JHEP} {\bf 03} (2014) 111,
  [\href{http://xxx.lanl.gov/abs/1304.7760}{{\tt arXiv:1304.7760}}].

\bibitem{Coriano:2022jkn}
C.~Corian\`o, M.~M. Maglio, and R.~Tommasi, {\it {Four-point functions of
  gravitons and conserved currents of CFT in momentum space: testing the
  nonlocal action with the TTJJ}},
  \href{http://xxx.lanl.gov/abs/2212.12779}{{\tt arXiv:2212.12779}}.

\bibitem{Coriano:2020ees}
C.~Corian\`o and M.~M. Maglio, {\it {Conformal field theory in momentum space
  and anomaly actions in gravity: The analysis of three- and four-point
  function}},  {\em Phys. Rept.} {\bf 952} (2022) 2198,
  [\href{http://xxx.lanl.gov/abs/2005.06873}{{\tt arXiv:2005.06873}}].

\bibitem{Marotta:2022jrp}
R.~Marotta, K.~Skenderis, and M.~Verma, {\it {Momentum space CFT correlators of
  non-conserved spinning operators}},
  \href{http://xxx.lanl.gov/abs/2212.13135}{{\tt arXiv:2212.13135}}.

\bibitem{Coriano:2023hts}
C.~Corian\`o, S.~Lionetti, and M.~M. Maglio, {\it {Parity-odd 3-point functions
  from CFT in momentum space and the chiral anomaly}},  {\em Eur. Phys. J. C}
  {\bf 83} (2023), no.~6 502, [\href{http://xxx.lanl.gov/abs/2303.10710}{{\tt
  arXiv:2303.10710}}].

\bibitem{Capper:1974ic}
D.~M. Capper and M.~J. Duff, {\it {Trace anomalies in dimensional
  regularization}},  {\em Nuovo Cim. A} {\bf 23} (1974) 173--183.

\bibitem{Christensen:1978gi}
S.~M.~Christensen and M.~J.~Duff,
{\it {Axial and Conformal Anomalies for Arbitrary Spin in Gravity and Supergravity}},
 {\em Phys. Lett. B} {\bf76} (1978), 571

\bibitem{Christensen:1978md}
S.~M.~Christensen and M.~J.~Duff,
{\it {New Gravitational Index Theorems and Supertheorems}},
{\em Nucl. Phys. B} {\bf 154} (1979), 301-342

\bibitem{Duff:2020dqb}
M.~J. Duff, {\it {Weyl, Pontryagin, Euler, Eguchi and Freund}},  {\em J. Phys.
  A} {\bf 53} (2020), no.~30 301001,
  [\href{http://xxx.lanl.gov/abs/2003.02688}{{\tt arXiv:2003.02688}}].

\bibitem{Deser:1980kc}
S.~Deser, M.~J. Duff, and C.~J. Isham, {\it {GRAVITATIONALLY INDUCED CP
  EFFECTS}},  {\em Phys. Lett. B} {\bf 93} (1980) 419--423.

\bibitem{Duff:1980qv}
M.~J. Duff and P.~van Nieuwenhuizen, {\it {Quantum Inequivalence of Different
  Field Representations}},  {\em Phys. Lett. B} {\bf 94} (1980) 179--182.

\bibitem{Sezgin:1980tp}
E.~Sezgin and P.~van Nieuwenhuizen, {\it {Renormalizability Properties of
  Antisymmetric Tensor Fields Coupled to Gravity}},  {\em Phys. Rev. D} {\bf
  22} (1980) 301.

\bibitem{Coriano:2023gxa}
C.~Corian\`o, S.~Lionetti, and M.~M. Maglio, {\it {Parity-violating CFT and the
  gravitational chiral anomaly}},  {\em Phys. Rev. D} {\bf 109} (2024), no.~4
  045004, [\href{http://xxx.lanl.gov/abs/2309.05374}{{\tt arXiv:2309.05374}}].

\bibitem{Coriano:2023cvf}
C.~Corian\`o, S.~Lionetti, and M.~M. Maglio, {\it {CFT Correlators and
  CP-Violating Trace Anomalies}},
  \href{http://xxx.lanl.gov/abs/2307.03038}{{\tt arXiv:2307.03038}}.

\bibitem{Coriano:2018bbe}
C.~Corian\`o and M.~M. Maglio, {\it {Exact Correlators from Conformal Ward
  Identities in Momentum Space and the Perturbative $TJJ$ Vertex}},  {\em Nucl.
  Phys.} {\bf B938} (2019) 440--522,
  [\href{http://xxx.lanl.gov/abs/1802.07675}{{\tt arXiv:1802.07675}}].

\bibitem{Coriano:2019nkw}
C.~Corian\`o, M.~M. Maglio, and D.~Theofilopoulos, {\it {Four-Point Functions
  in Momentum Space: Conformal Ward Identities in the Scalar/Tensor case}},
  \href{http://xxx.lanl.gov/abs/1912.01907}{{\tt arXiv:1912.01907}}.

\bibitem{Edgar:2001vv}
S.~B. Edgar and A.~Hoglund, {\it {Dimensionally dependent tensor identities by
  double antisymmetrization}},  {\em J. Math. Phys.} {\bf 43} (2002) 659--677,
  [\href{http://xxx.lanl.gov/abs/gr-qc/0105066}{{\tt gr-qc/0105066}}].

\bibitem{lovelock_1970}
D.~Lovelock, {\it Dimensionally dependent identities},  {\em Mathematical
  Proceedings of the Cambridge Philosophical Society} {\bf 68} (1970), no.~2
  345?350.

\bibitem{Bzowski:2017poo}
A.~Bzowski, P.~McFadden, and K.~Skenderis, {\it {Renormalised 3-point functions
  of stress tensors and conserved currents in CFT}},
  \href{http://xxx.lanl.gov/abs/1711.09105}{{\tt arXiv:1711.09105}}.

\bibitem{Barvinsky:1995it}
A.~O. Barvinsky, A.~G. Mirzabekian, and V.~V. Zhytnikov, {\it {Conformal
  decomposition of the effective action and covariant curvature expansion}},
  in {\em {6th Moscow Quantum Gravity}}, 6, 1995.
\newblock \href{http://xxx.lanl.gov/abs/gr-qc/9510037}{{\tt gr-qc/9510037}}.

\bibitem{Fradkin:1978yw}
E.~S. Fradkin and G.~A. Vilkovisky, {\it {Conformal Off Mass Shell Extension
  and Elimination of Conformal Anomalies in Quantum Gravity}},  {\em Phys.
  Lett. B} {\bf 73} (1978) 209--213.

\bibitem{Riegert:1984kt}
R.~J. Riegert, {\it {A Nonlocal Action for the Trace Anomaly}},  {\em Phys.
  Lett.} {\bf 134B} (1984) 56--60.

\bibitem{Capozziello:2024qol}
S.~Capozziello, A.~Mazumdar, and G.~Meluccio, {\it {Can nonlocal gravity really
  explain dark energy?}},  \href{http://xxx.lanl.gov/abs/2403.11301}{{\tt
  arXiv:2403.11301}}.

\bibitem{Capozziello:2021krv}
S.~Capozziello and F.~Bajardi, {\it {Nonlocal gravity cosmology: An overview}},
   {\em Int. J. Mod. Phys. D} {\bf 31} (2022), no.~06 2230009,
  [\href{http://xxx.lanl.gov/abs/2201.04512}{{\tt arXiv:2201.04512}}].

\bibitem{Maggiore:2014sia}
M.~Maggiore and M.~Mancarella, {\it {Nonlocal gravity and dark energy}},  {\em
  Phys. Rev. D} {\bf 90} (2014), no.~2 023005,
  [\href{http://xxx.lanl.gov/abs/1402.0448}{{\tt arXiv:1402.0448}}].

\end{thebibliography}

\end{document}